\documentclass[apj, letterpaper]{emulateapj}  % for apj-lookalike
\pdfoutput=1

\usepackage{amsmath}
\usepackage{lineno}
\begin{document}

\slugcomment{\bf}
\slugcomment{published in ApJ, 804:60}

\title{Atmospheric dynamics of terrestrial exoplanets over a wide range of orbital and atmospheric parameters}

\shorttitle{Atmospheric dynamics of terrestrial exoplanets}
\shortauthors{Kaspi and Showman}

\author{ Yohai Kaspi\altaffilmark{1} Adam P. Showman\altaffilmark{2},}

\altaffiltext{1}{Department of Earth and Planetary Sciences, Weizmann
  Institute of Science, 234 Herzl st., 76100, Rehovot, Israel; yohai.kaspi@weizmann.ac.il}
\altaffiltext{2}{Department of Planetary Sciences and Lunar and Planetary
Laboratory, The University of Arizona, 1629 University Blvd., Tucson, AZ 85721 USA}

% Abstract 
\begin{abstract}
\label{Abstract}

The recent discoveries of terrestrial exoplanets and super Earths
extending over a broad range of orbital and physical parameters suggests
that these planets will span a wide range of climatic regimes. Characterization
of the atmospheres of warm super Earths has already begun and will
be extended to smaller and more distant planets over the coming decade.
The habitability of these worlds may be strongly affected by their
three-dimensional atmospheric circulation regimes, since the global
climate feedbacks that control the inner and outer edges of the habitable
zone---including transitions to Snowball-like states and runaway-greenhouse
feedbacks---depend on the equator-to-pole temperature differences,
patterns of relative humidity, and other aspects of the dynamics. Here,
using an idealized moist atmospheric general circulation model (GCM)
including a hydrological cycle, we study the dynamical principles
governing the atmospheric dynamics on such planets. We show how the
planetary rotation rate, planetary radius, surface gravity, stellar flux, optical thickness and atmospheric mass affect the 
atmospheric circulation and temperature distribution on such planets. 
Our simulations demonstrate that equator-to-pole temperature differences,
meridional heat transport rates, structure and strength of the winds, and
the hydrological cycle vary strongly with these parameters, implying
that the sensitivity of the planet to global climate feedbacks will
depend significantly on the atmospheric circulation. We elucidate the possible
climatic regimes and diagnose the mechanisms controlling the formation
of atmospheric jet stream, Hadley and Ferrel cells and latitudinal
temperature differences. Finally, we discuss the implications for
understanding how the atmospheric circulation influences the global climate.
\end{abstract}

\section{Introduction\label{sec:Introduction}}

Since the mid-1990s, nearly 2000 planets have been discovered
around other stars. The first to be discovered were giant planets
with short orbital periods, and since then, many smaller planets with
longer orbital periods have been identified. The planets can be generally
divided into two types: planets close to their parent star that
become synchronously locked, resulting in one side constantly being
heated from their parent star, and asynchronously rotating planets
with a diurnal cycle --- similar to Earth and most solar system planets.
In this study we focus on the latter type, and within these, we focus
on terrestrial planets, i.e., those in which the atmospheric dynamics
are limited to a thin spherical shell overlying a solid surface. These
planets span a large range of masses, radii, densities, incident stellar
fluxes, orbital periods, and orbital eccentricities. The goal of this
study is to characterize the range of possible climatic regimes these
planets might encompass, and characterize how the climate and habitability
depend on these orbital, planetary, and atmospheric parameters.

Although exoplanet discovery and characterization began with giant
planets, emphasis is gradually shifting to smaller worlds. Approximately
100 planets with masses less than $\sim10$ Earth masses have been
discovered\footnote{www.exoplanet.eu}, with many hundreds of additional
candidates identified by the NASA Kepler spacecraft
\citep{Borucki2011}. Planets toward the upper end of this mass range
may typically constitute mini Neptunes with no solid surfaces (e.g.,
\citealp{Valencia2007,Adams2008,Rogers2011,Nettelmann2011,Fortney2013}),
but planets toward the lower end are more likely terrestrial planets
with solid surfaces and relatively thin atmospheres. Importantly,
discoveries to date include a number of planets with masses and/or
radii less than those of Earth (e.g.,
\citealp{Fressin2012,Muirhead2012,Borucki2013,Barclay2013} see review
by \citealp{Sinukoff2013}), as well as numerous planets $\sim1-3$
Earth radii in size. This overall population of super Earths and
terrestrial planets includes not only hot, inhabitable objects blasted
by starlight (CoRoT-7b and Kepler-10b being prominent examples;
\citealp{Leger2011,Batalha2011}), but also many planets receiving
$\sim0.2$ to several times the incident stellar flux Earth receives
from the Sun, with effective temperatures of $\sim200-400$~K
(\citealp{Muirhead2012,Dressing2013, Quintana2014}; see Figure 7 in
\citealp{Ballard2013} for a visual summary). Depending on atmospheric
composition, these moderate stellar fluxes put these planets in or
near the classical habitable zones around their stars.

Atmospheric characterization of super Earths, while difficult, has
already begun. Attention to date has focused on GJ 1214b, a 6.5-Earth-mass,
2.7-Earth-radius super Earth orbiting a nearby M dwarf \citep{Charbonneau2009}.
Transit spectroscopy in visible and near-infrared (IR) wavelengths
indicates a relatively flat spectrum, ruling out hydrogen-dominated,
cloud-free atmospheres and favoring instead a high-molecular-weight
(e.g., water-dominated) atmosphere and/or the presence of clouds that
obscure spectral features (e.g., \citealp{Bean2010,Bean2011,Desert2011,Berta2012,DeMooij2012,Fraine2013,Teske2013, Kreidberg2014}).
The secondary eclipses of this relatively cool ($\sim500$~K) planet
have recently been detected by Spitzer \citep{Gillon2014}. The NASA
Transiting Exoplanet Survey Satellite (TESS, to be launched in 2017)
and ESA's Planetary Transits and Oscillations of 
Stars satellite (PLATO, to be launched in 2024)
will search for additional, observationally favorable super Earths,
and upcoming platforms including the James Webb Space Telescope
(JWST), the Characterizing Exoplanets Satellite (CHEOPS), and
the groundbased Thirty Meter Telescope (TMT) and the European
Extremely Large Telescope (E-ELT) will be capable of
characterizing their atmospheres. These developments indicate that,
in the coming decade, observational techniques currently used to characterize
the atmospheres of hot Jupiters will be applied to super Earths and
terrestrial planets, placing constraints on their atmospheric composition,
thermal structure, and climate. In principle, visible and infrared
light curves, ingress/egress mapping during secondary eclipse, and
shapes of spectral lines during transit could lead to constraints
on longitudinal temperature variations, latitudinal temperature variations,
cloud patterns, and vertical temperature profiles.

These observational developments provide a strong motivation for
investigating the possible atmospheric circulation regimes of
terrestrial exoplanets over a wide range of conditions. Such an
investigation---as we carry out here---can provide a theoretical
framework for interpreting future measurements of these planets and
assessing their habitability. Moreover, such an effort can help to
answer the fundamental, unsolved theoretical question of how the
atmospheric circulations of terrestrial planets---broadly
defined---vary with incident stellar flux, atmospheric mass,
atmospheric opacity, planetary rotation rate, gravity, and other
parameters. While general circulation model (GCM) studies of Mars,
Venus, Titan, and especially Earth provide insights into the relevant
dynamical mechanisms, these models are typically constructed for a
narrow range of conditions specific to those planets, and thereby
provide only limited understanding of how the circulation regimes vary
across the continuum of possible conditions. More recently, several
authors have carried out GCM studies of terrestrial exoplanets,
emphasizing synchronously locked, slowly rotating planets
(\citealp{Joshi1997,Joshi2003,Merlis2010,Heng2011,Wordsworth2011,Selsis2011,
Yang2013, Yang2014, Hu2014}). Despite the insights provided by these studies,
only a small subset of possible conditions have yet been explored,
especially for planets that are not synchronously rotating. It thus
remains unclear, for example, how the equator-pole temperature
differences, vertical temperature profiles, wind speeds, and
properties of the Hadley cell, jet streams, instabilities, and waves
should vary with the atmospheric mass, atmospheric opacity, planetary
rotation rate, and other parameters.

In addition to its inherent interest, knowledge of how the atmospheric
circulation varies over a broad range of parameters will aid an
understanding of how the circulation interacts with global-scale
climate feedbacks to control planetary habitability. For example, the
conditions under which planets enter globally glaciated or runaway
greenhouse states depend on the equator-to-pole temperature
difference, the 3D distribution of humidity, and other aspects of the
circulation (e.g.,
\citealp{Voigt2011, Pierrehumbert2011, Leconte2013, Wordsworth2013, Yang2013, Forget2014}). See \citet{Showman2014} for a review of our current
understanding of the atmospheric circulation of terrestrial
exoplanets.

In this study we focus on the leading order mechanisms controlling the
general circulation. We refer to the general circulation
\citep{Lorenz1967} as the longitudinally averaged circulation of the
atmosphere, assuming there are no longitudinal asymmetries on the
planets (e.g., continents, topography). This allows us to keep the
analysis as simple as possible while maintaining the leading-order
forcing of the climate system.  While longitudinal asymmetries in the
surface can modify the circulation, this is often a second-order
effect; for example, even on Earth which has significant longitudinal
and hemispherical differences in continental distribution, the leading
order climate is zonally symmetric. As will be discussed, for Earth
this is a consequence of the rapid planetary rotation, but even on
slower rotating planets or moons (e.g., Titan) similar hemispherical
asymmetry is found to leading order (e.g.,
\citealp{Mitchell2006,Aharonson2009,Schneider2012}).

%Mapping the vast space of possible orbital and atmospheric parameters
%is a task that the field will need to undertake in the coming
%years. Here, as a first attempt at this, we begin with focuing on
%several parameters, and map them one-by-one, while leavving the other
%parameters fixed at the reference climate. 

% climatic regimes, re
%As the parameter space of possible orbital and
%atmospheric parameters is extremely
%large, and as a first attempt to explore a parameter space relevant to
%freely rotating planets much wider
%than what we find in the solar system, in this study we preform a p

As the parameter space of possible orbital and atmospheric parameters is extremely
large, in this study we focus on several parameters we
identify as important for understanding the dynamics of
freely-rotating terrestrial atmospheres. Specifically, we investigate how
the general circulation is affected by rotation rate,
atmospheric mass, stellar flux, surface gravity, optical thickness and
planetary mass over a wide range (much broader than that of the solar system planets). Although
even this subset of parameters are tied together and affect one another (e.g.,
varying atmospheric mass affects optical thickness), we study them
one-by-one in attempt to capture the effect of each one of them
separately on the general circulation. This can provide a baseline for more detailed studies and
multi-dimensional explorations of the vast parameter space in attempt
to simulate more realistic scenarios. For simplicity, we perform all
experiments at perpetual equinox conditions, ignoring effects of
obliquity and eccentricity and therefore seasonal
variations. We also ignore other
important components of the climate system such as clouds, ice,
sea-ice and albedo variations. Thus, although the model itself is fully 3D, the forcing (i.e.,
the imposed starlight) is both north-south hemispherically symmetric and
zonally symmetric, and includes only the
most basic components of the climate system. Yet, it still captures the
most fundamental features of the general circulation (e.g., jets, waves, Hadley
and Ferrel cells), and gives a wide range dynamical phenomena which can be compared
against the Solar System terrestrial planetary atmospheres.

%As the parameter space of possible orbital and atmospheric parameters is extremely
%large it is impossible to map all possible climatic regime of this
%parameter space. Therefore, this paper comes as a first attempt to
%address this in the context of terrestrial freely-rotating exoplanets
%in order to provide a framework for more detailed studies. We
%therefore focus on six important parameters, namely the rotation rate,
%atmospheric mass, stellar flux, surface gravity, thermal opacity and
%planetary mass and explore them one-by-one in compared to a reference
%climate. Although,  For simplicity, we perform all
%experiments at perpetual equinox conditions, ignoring effects of
%obliquity and eccentricity and therefore seasonal variations, meaning
%the simulations are all hemispherically symmetric 
%hemispherical asymmetry to the climate system.

Section 2 introduces our model and presents control
simulations for modern-Earth conditions, which provides a reference
against which to compare our parameter variations. Section 3
describes our main results, where every subsection presents a separate
set of simulations where the dependence on one parameter compared to
the reference climate is explored. For each parameter study, we present
detailed description of the general circulation, particularly focusing
on the equator-to-pole temperature difference, the poleward heat transports,
Hadley and Ferrell cells and the jet streams. Section 4
concludes and summarizes implications for observables and habitability.

\section{Model\label{sec:Model}}

\subsection{Model description\label{sub:Model-description}}

To explore the sensitivity of the general circulation to the basic
characteristics of the planet we use an idealized General Circulation
Model (GCM). The idealized GCM is based on the Flexible Modeling System
(FMS) of NOAA's Geophysical Fluid Dynamics Laboratory (GFDL) \citep{Anderson2004,Held1994}.
It is a three-dimensional model of a spherical aquaplanet\footnote{An
  aquaplanet is defined as a terrestrial planet with a (relatively)
  thin atmosphere, a global liquid-water ocean with no continents, and
  a discrete interface between the ocean and atmosphere; this is to be
  distinguished from the term ``ocean planet'' \citep{Leger2004} which
  sometimes is used to denote fully fluid planets (mini-Neptunes)
  composed predominantly of H$_2$O.}, at perpetual equinox and solves
the primitive equations for fluid motion on the sphere for an ideal-gas
atmosphere in the reference frame of the rotating planet with rotation
rate $\Omega$. The primitive equations in spherical coordinates are
given, using pressure as a vertical coordinate, by
\begin{eqnarray}
\frac{Du}{Dt}-2\Omega v \sin\theta-\frac{uv}{a}\tan\theta & = & -\frac{1}{a\cos\theta}\frac{\partial\Phi}{\partial\lambda}-\Sigma_{u},\label{eq:Du}\\
\frac{Dv}{Dt}+2\Omega u \sin\theta+\frac{u^{2}}{a}\tan\theta & = & -\frac{1}{a}\frac{\partial\Phi}{\partial\theta}-\Sigma_{v},\label{eq:Dv}\\
0 & = & -\frac{\partial\Phi}{\partial\ln p}-R_{d}T_{v},\label{eq:Dw}\\
\nabla\cdot\mathbf{u} & = & 0,\label{eq:mass}\\
\frac{DT}{Dt}-\frac{R_{d}T_v\omega}{c_{p}p} & = & Q_{r}+Q_{c}+Q_{b},\label{eq:DT}\\
 &  & \text{}\nonumber 
\end{eqnarray}
where $u$, $v,$ and $\omega$ are the longitudinal
$\left(\lambda\right)$, latitudinal $\left(\theta\right)$ and pressure
$\left(p\right)$ velocities respectively, $\Phi$ is geopotential, $T$
is temperature and $T_{v}$ is the virtual temperature\footnote{The
  virtual temperature is defined as $T_{v}=T\left(1+\left(m_d/m-1\right)q\right)$, where $m_d$ and
  $m$ are the mean molecular mass of dry and moist air, respectively, and $q$ is the specific humidity;
  e.g., \citet{Bohren1998}.  Virtual temperature is the temperature
  that air at a given pressure and density would have if the air were
  completely free of water vapor; and thus, always equal or greater
  than the temperature.}.
The material derivative is given by
$\frac{D}{Dt}=\frac{\partial}{\partial t}+\mathbf{u}\cdot\nabla,$
where $\mathbf{u}=\left(u,v,\omega\right)$ and $t$ is
time. $\Sigma_{u}$, $\Sigma_{v}$ are the surface stress terms from the
boundary layer (see below), $R_{d}=287$ J~kg$^{-1}$~K$^{-1}$ is the
dry gas constant for air, $c_{p}=1004$ J~kg$^{-1}$~K$^{-1}$ is the
specific heat of air, $a$ is the planetary radius and $Q_{r}$, $Q_{c}$
and $Q_{b}$ are the radiative, convective and boundary layer heating
per unit mass, respectively (see
below). In Eq.~(\ref{eq:Du}--\ref{eq:Dw}) we have made the traditional
assumptions for terrestrial atmospheres that due to the shallowness of
the atmosphere compared to the radius of the planet the horizontal
Coriolis terms and some of the metric terms are negligible
\citep{Vallis2006}.  Note that, because pressure is the vertical
coordinate, Eq.~(\ref{eq:mass}) does {\it not} imply that density is
constant; indeed, we use the ideal-gas equation of state, which allows
significant density variations vertically and horizontally.

The primitive equations are solved in
vorticity-divergence form using the spectral transform method in the
horizontal and finite differences in the vertical \citep{Bourke1974}.
We mostly use a horizontal resolution of 42 spectral modes (T42),
which corresponds to about $2.8^{\circ}\times2.8^{\circ}$ resolution
in longitude and latitude, but for cases where eddy length scales become
small (e.g., high rotation rates) we increase the horizontal
resolution up to T170 ($0.7^{\circ}\times0.7^{\circ}$). The vertical
coordinate is $\sigma=p/p_{s}$ (pressure $p$ normalized by surface
pressure $p_{s}$). It is discretized with 30 levels, unequally spaced
to ensure adequate resolution in the lower troposphere and near the
tropopause.  The uppermost full model level has a mean pressure of
$0.46\%$ of the mean surface pressure. For simplicity, the model does not include clouds,
continental effects, sea-ice or snow. All simulations have been spun
up to statistically steady state for at least 1500 simulation days,
and the results presented here have been then averaged over at least
the subsequent 1500 simulation days, and are all in a statistically
stable state.

\subsubsection{Radiative transfer\label{sub:Radiative-transfer}}

\begin{figure*}
\begin{centering}
\includegraphics[scale=0.4]{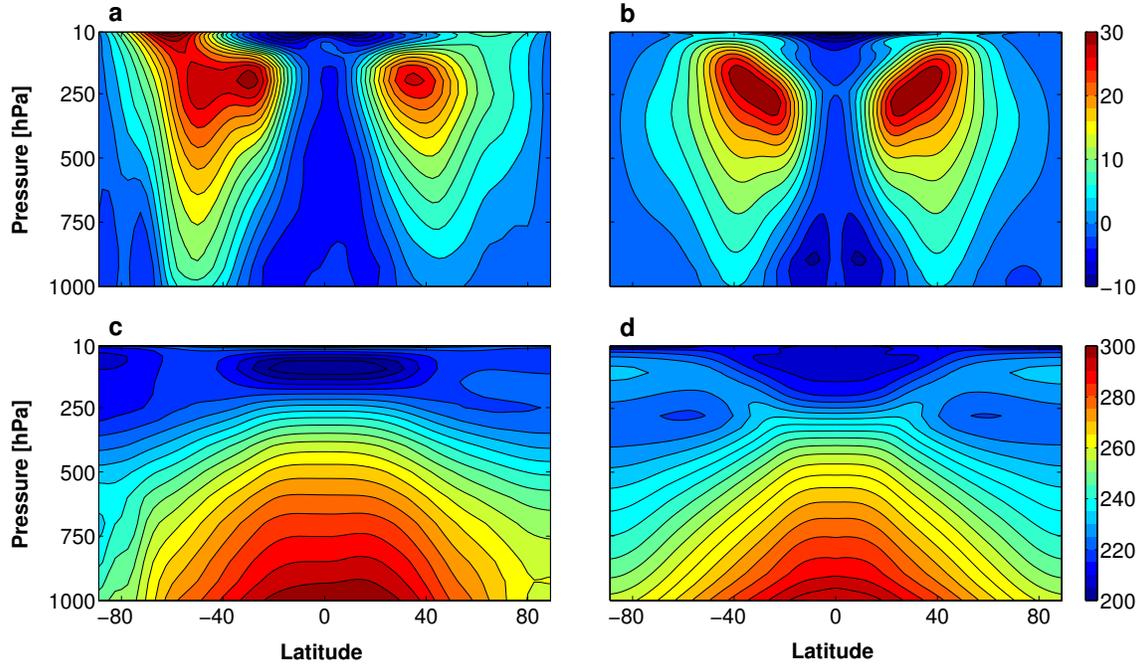}
\par\end{centering}

\caption{\label{fig:compare_ncep_model} The zonal mean zonal wind (m~s$^{-1}$)
(top), and zonal mean temperature distribution (K) (bottom) from
Earth's NCEP reanalysis data averaged over the years 1970-2012 (left) and the reference
simulation (right). }
\end{figure*}

Radiative transfer is represented by a standard two-stream gray radiation
scheme \citep{Held1982,Frierson2006} given by 
\begin{eqnarray}
\frac{dU}{d\tau} & = & U-\sigma_{_{\rm{SB}}}T^{4},\label{eq:DU}\\
\frac{dD}{d\tau} & = & \sigma_{_{\rm{SB}}}T^{4}-D,\label{eq:DD}
\end{eqnarray}
where $U$ is the longwave\footnote{Here, longwave refers to the long-wavelength
infrared radiation associated with planetary emission; this is to be distinguished
from shortwave radiation which represents the stellar flux.} upward flux and $D$ is the longwave downward flux, with
the boundary condition at the surface being $U\left(\tau\left(z=0\right)\right)=\sigma_{_{\rm{SB}}}T_{s}^{4}$
, where $T_{s}$ is the surface temperature (see below), and $D\left(\tau=0\right)=0$
at the top of the atmosphere. $\sigma_{_{\rm{SB}}}=5.6734\times10^{-8}$ ~W~m$^{-2}$~K$^{-4}$
is the Stefan-Boltzmann constant. The longwave optical thickness $\tau$
is given by 
\begin{eqnarray}
\tau & = & \left[f_{l}\sigma+\left(1-f_{l}\right)\sigma^{4}\right]\left[\tau_{e}+\left(\tau_{p}-\tau_{e}\right)\sin^{2}\theta\right],\label{eq:tau}
\end{eqnarray}
where $f_{l}$, $\tau_{e}$, and $\tau_{p}$ are constants; this implies
that the longwave and shortwave optical depths only depend
on latitude and pressure. The longwave optical thickness at the equator
and pole $\tau_{e}=8.4$ and $\tau_{p}=2.2$ respectively, and $f_{l}=0.2$,
are chosen to mimic roughly an Earth like equinoctial meridional and
vertical temperature distribution (Fig.~\ref{fig:compare_ncep_model}).
The quartic term in (\ref{eq:tau}) represents the rapid increase
of opacity near the surface due to water vapor under Earth-like conditions
\citep{Frierson2006}. Note that, because the constants in 
Eq.~(\ref{eq:tau}) are specified, our experiments do not include the
water-vapor feedback in which variations in water vapor cause 
variations in opacity.  The radiative source term in
the atmospheric interior is then given by 
\begin{eqnarray}
Q_{r} & = & \frac{g}{c_{p}}\frac{\partial}{\partial p}\left(U-D-R_S\right),\label{eq:Qr}
\end{eqnarray}
where insolation is imposed equally between hemispheres with the
insolation $R_{s}$ set as 
\begin{eqnarray}
R_{s} & = & \frac{S_{0}}{4}\left[1+\frac{\Delta_{s}}{4}\left(1-3\sin^{2}\theta\right)\right]e^{-\tau_{s}\sigma^{2}},\label{eq:Rs}
\end{eqnarray}
where $S_{0}=1360$ W~m$^{-2}$, $\Delta_{s}=1.2$, and the parameter
$\tau_{s}=0.08$ controls the vertical absorption of solar radiation
in the atmosphere. These parameters have also been set to mimic an
Earth-like climate (Fig.~\ref{fig:compare_ncep_model}).

\subsubsection{Surface boundary layer\label{sub:Surface-boundary-layer}}

Our surface boundary-layer scheme is similar to that of \citet{Frierson2006}.
The lower boundary of the GCM is a uniform water covered slab, with
an albedo of $\alpha=0.35$. A planetary boundary layer scheme with
Monin-Obhukov surface fluxes \citep{Obukhov1971}, which depend on
the stability of the boundary layer, links atmospheric dynamics to
surface fluxes of momentum, latent heat, and sensible heat. The roughness
length for momentum is $5\times10^{-3}$, and for moisture and heat fluxes is $1\times10^{-5}$~m,
and an additive gustiness term of 1~m~s$^{-1}$ in surface velocities comes
to represent subgrid-scale wind fluctuations. These values yield 
energy fluxes and a climate similar to Earth\textquoteright{}s in
the aquaplanet setting of our simulations (Fig.~\ref{fig:compare_ncep_model}).
Our results are not very sensitive to the choice of these parameters.

The lower boundary is a ``slab ocean'', comprising a vertically
uniform layer of liquid water of depth $H$, here taken to be 1~m thick, with no dynamics
and a horizontally varying temperature evolving as
\begin{eqnarray}
c_{po}\rho_0H\frac{\partial T_{s}}{\partial t} & = & R_{s}\left(1-\alpha\right)-R_{u}-L_{e}E-S,\label{eq:DTs}
\end{eqnarray}
where $c_{po}=3989$~J~kg$^{-1}$~K$^{-1}$ is the surface heat capacity, $\rho=1035$~kg~m$^{-3}$ is the effective
density of the slab ocean, $L_{e}=2.5\times10^{6}$
J~kg$^{-1}$ is the latent heat of vaporization and $R_{u}$ is the
net upward longwave flux from the ocean surface. $E$ and $S$ and $\Sigma$ are
the surface evaporative heat flux, sensible heat
flux\footnote{Sensible heat refers to 
energy stored, i.e., thermal energy (as distinct from latent heat).  Here, the sensible heat
flux is the conductive heat flux from the slab ocean to the atmosphere.} and surface
stress respectively which are given by
\begin{eqnarray}
E & = & \rho_{a}C\left|\mathbf{u}_{a}\right|\left(q_{a}-q_{s}^{*}\right),\label{eq:evaporative heating}\\
S & = &\rho_{a}c_{p}C\left|\mathbf{u}_{a}\right|\left(T_{a}-T_{s}\right),\label{eq:sensible heating}\\
\mathbf{\Sigma} & = & C\rho_{a}\left|\mathbf{u}_{a}\right|\left|\mathbf{u}_{a}\right|,\label{eq: surface stress}
\end{eqnarray}
where $\rho_{a},\mathbf{u}_{a},T_{a}$ and $q_{a}$ are the density,
horizontal wind, temperature and humidity at the lowest atmospheric
level. $q_{a}^{*}$ is the saturation specific humidity at the lowest
model layer (see below), and $C$ is a drag coefficient which decays
with height following Monin-Obhukov theory \citep{Obukhov1971,Frierson2006}.
For further details see \citet{Anderson2004}. The neglect of horizontal mixing and heat transport in the ocean
is a reasonable assumption in the Earth climate regime, since only
one-third of Earth's meridional heat transport occurs in the ocean, with
the remaining two-thirds occurring in the atmosphere.  Nevertheless, there
may be some exoplanet regimes in which this assumption
breaks down (e.g., \citealt{Hu2014}).

Above the boundary layer, horizontal $\nabla^{8}$ hyperdiffusion
in the vorticity, divergence, and temperature equations is the only
frictional process. The hyperdiffusion coefficient is chosen to give
a damping time scale of $12$~hrs at the smallest resolved scale.

\subsubsection{Hydrological cycle\label{sub:Moist-thermodynamics}}

We include a hydrological cycle involving the evaporation, condensation,
and transport of water vapor.
Moisture is calculated at every grid point and depends on the surface
evaporative fluxes and the convection giving the moisture equation
\begin{eqnarray}
\frac{Dq}{Dt} & = & g\frac{\partial E}{\partial p}-\frac{Q_{c}}{L_{e}}.\label{eq:Dq}
\end{eqnarray}
A large-scale (grid-scale)
condensation scheme ensures that the mean relative humidity in a grid
cell does not exceed 100\% \citep{Frierson2007a,O'Gorman2008a}. Only the vapor-liquid phase change is
considered, and the saturation vapor pressure $e_{s}$ is calculated
from a simplified Clausius-Clapeyron relation given by 
\begin{eqnarray}
e_{s}\left(T\right) & = & e_{0}\exp\left[-\frac{L_{e}}{R}\left(\frac{1}{T}-\frac{1}{T_{0}}\right)\right],\label{eq:Clausius-Clapiron}
\end{eqnarray}
where $R_{v}=461.5$ J~kg$^{-1}$K$^{-1}$ is the gas constant for
water vapor, $e_{0}=610.78$ Pa and $T_{0}=$273.16 K. The saturation
specific humidity is then calculated by $q_{s}=\frac{R_{v}e_{s}}{R_{d}p}$.
Moist convection is represented by a Betts-Miller like quasi-equilibrium convection
scheme \citep{Betts1986a,Betts1986b}, relaxing temperatures toward a moist adiabat with a time scale
of 2 hours, and water vapor toward a profile with fixed relative humidity
of 70\% relative to the moist adiabat, whenever a parcel lifted from
the lowest model level is convectively unstable. Large-scale condensation
removes water vapor from the atmosphere when the specific humidity
on the grid scale exceeds the saturation.

\subsection{Reference climate\label{sub:Reference-climate}}

Before presenting the dependence of the climate on the orbital and
atmospheric parameters, we begin by presenting below the reference
climate against which all experiments are compared. We choose
this to be a climate similar to Earth's climate,  which to leading
order is a well-understood
regime, and allows us to check the validity of our model by comparing it
to observations.   Our strategy is to perform systematic parameter space
sweeps all in reference to this single reference climate. With the parameter
choice presented above the model reference climate is set to represent
an Earth like annual-mean climate. Fig.~\ref{fig:compare_ncep_model}
shows the zonally averaged temperature and wind fields for this reference
climate (right), and the annually averaged climate for Earth from the National
Centers for Environmental Prediction (NCEP) reanalysis data averaged
over 40 years (1970-2010, left). Despite the simplicity of the model the
resulting mean climate represents well Earth's annual mean climate.
The tropics are dominated by Hadley cells reaching nearly latitude
$30^{\circ}$ with a generally weak westward zonal flow, and midlatitudes
are dominated by an eastward jet streams with wind velocities reaching
about $30$~m~s$^{-1}$ at latitude $45^{\circ}$ and 200~hPa.
Of course the model results are hemispherically symmetric, in contrast
to the Earth data which varies considerably between hemispheres. In
addition, our model does not have a seasonal cycle, which affects
the climate despite not appearing directly in the annual averaged
climate. Despite these differences, the choice of parameters given
in section~\ref{sub:Model-description} yields a mean climate 
similar to observations. Fig.~\ref{fig:4 surface temp} (upper right)
shows the surface temperature map resembling that of Earth's observations
with wave number five Rossby waves dominating the midlatitude climate. 

\begin{figure}
\begin{centering}
\includegraphics[scale=0.45]{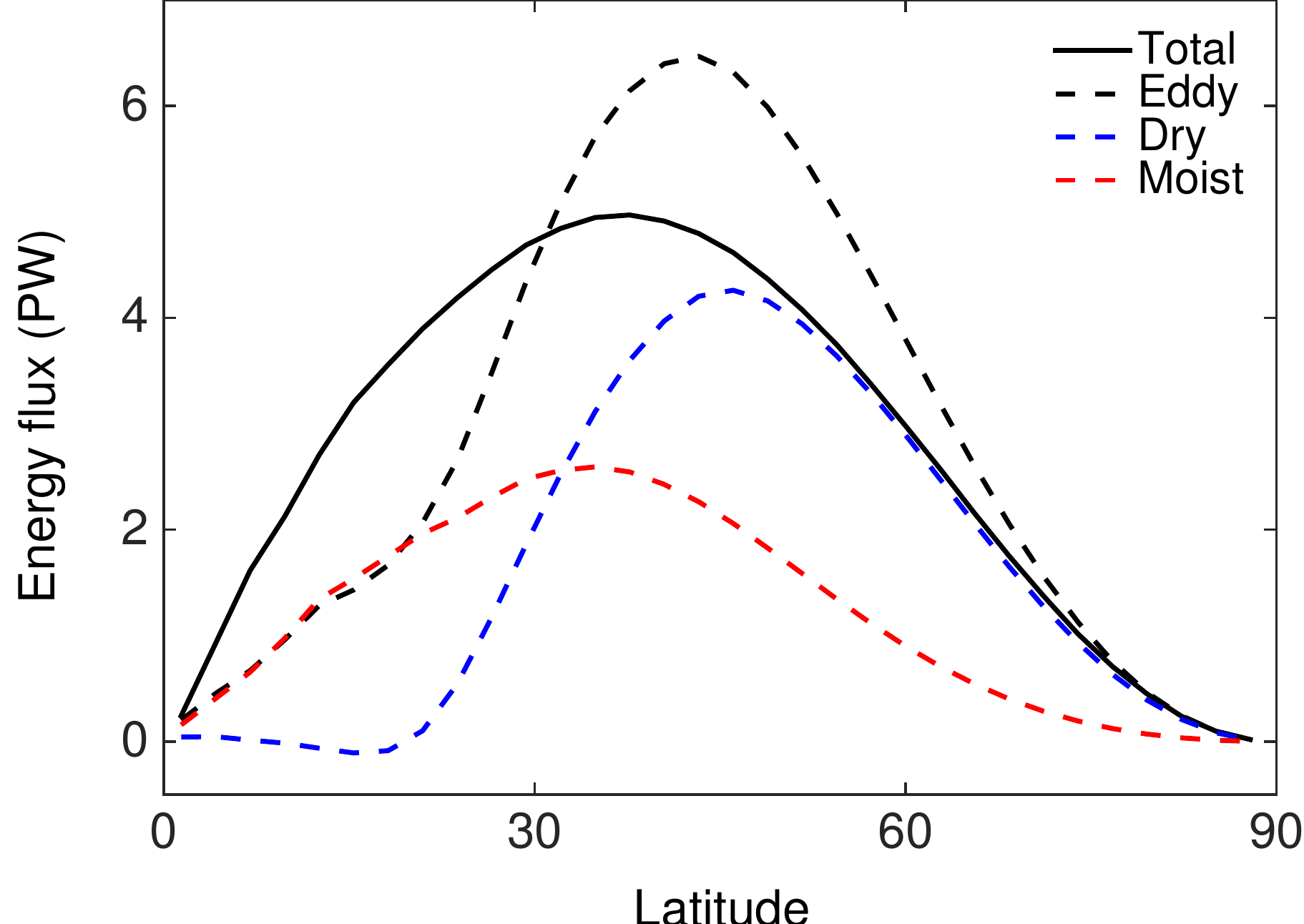}
\par\end{centering}

\caption{\label{fig:mse_flux} The total (full) and eddy (dashed) poleward
moist static energy flux (PW). The eddy flux is divided to its dry
(blue) and moist (red) components. }
\end{figure}

\begin{figure*}
\begin{centering}
\includegraphics[scale=0.35]{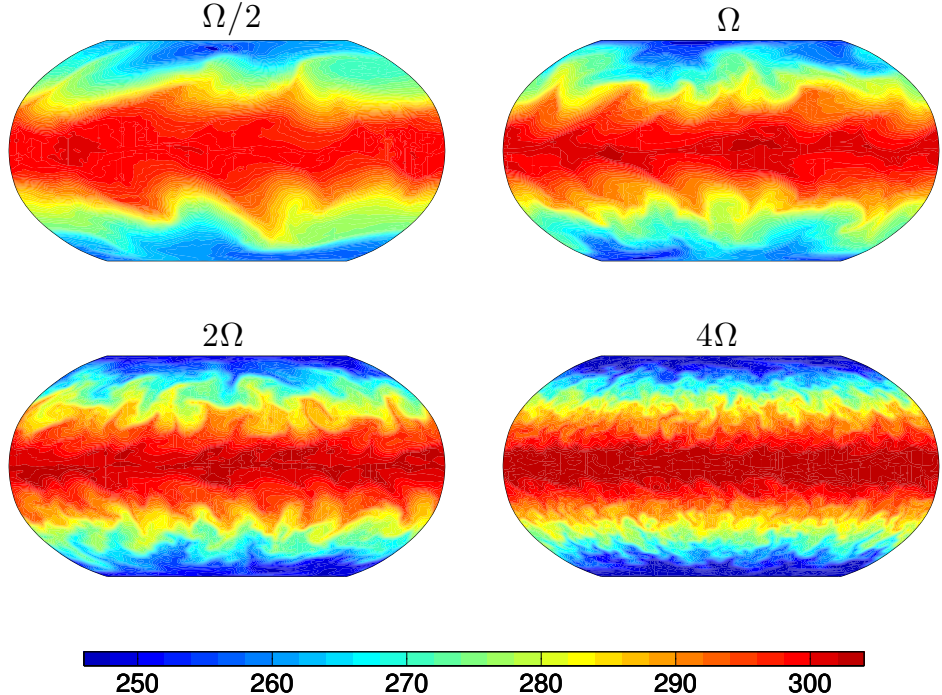}
\par\end{centering}

\caption{\label{fig:4 surface temp}Surface temperature (colorscale, in K)
illustrating the dependence of temperature and eddy scale on rotation
rate. Simulations have rotation rates ranging from half (upper left)
to four times that of Earth (lower right). Baroclinic instabilities
dominate the dynamics in mid- and high-latitudes, leading to baroclinic
eddies whose length scales decrease with increasing planetary rotation
rate.}
\end{figure*}

A main focus of this paper will be on what controls the equator-to-pole
temperature difference on terrestrial exoplanets. For Earth, if the
planet had no atmospheric and oceanic dynamics and the planet were
in pure radiative equilibrium, the equator-to-pole temperature difference
would be much larger than the actual value \citep{Hartmann1994}. The existence of dynamics
in the atmosphere, and particularly the fact that the atmosphere is
turbulent, leads to a net poleward heat flux, which results
in the cooling of the lower latitudes and heating of high latitudes,
and therefore a much more equable climate than a planet in pure radiative
equilibrium. This results in the reduction of the mean equator-to-pole
temperature difference. The total heat transport can be described
in terms of the moist static energy (MSE) defined as 
\begin{eqnarray}
m & = & c_{p}T+gz+L_{e}q,\label{eq:mse}
\end{eqnarray}
where all symbols were defined in section~\ref{sub:Model-description}.
For Earth's climate the global moist static energy flux is poleward
in both hemispheres and peaks at about $5\times10^{15}$~W in midlatitudes
(Fig.~\ref{fig:mse_flux}). The total flux can be divided into a time
mean component and a variation from the time mean coming from the
turbulence in the atmosphere. If we divide the meridional velocity
into a time mean denoted by an over-line and deviations from the time mean (eddies) with a prime
so that $v=\overline{v}+v'$, and do similarly for the moist static
energy then
\begin{eqnarray}
\overline{vm} & = & \overline{v}\,\overline{m}+\overline{v'm'}.\label{eq:mse_flux}
\end{eqnarray}
Fig.~\ref{fig:mse_flux} shows the zonal averaged total flux ($\overline{vm}$) and
the eddy flux ($\overline{v'm'}$), divided into the dry static energy
component $\overline{v's'}$, where $s=c_{p}T+gz$, and the latent
energy component $L_{e}\overline{v'q'}$. It shows the dominance of
the eddy term within the total flux (in midlatitudes it is even bigger
since the mean component is negative), and the fact that the dry and
moist components contribute roughly equally to this flux. For Earth,
the moist component is more dominant in low latitudes and the dry
component in high latitudes. This reflects the strong nonlinear dependence
of the saturation vapor pressure on temperature (Eq.~\ref{eq:Clausius-Clapiron}),
meaning that warmer climates will have stronger heat transport due
to latent heating (see
section~\ref{sub:Dependence-on-distance}). Since the surface
temperature is coupled to the atmosphere the radiation and evaporation also respond
to the MSE transport giving important feedbacks to the steady state
temperature through the Planck, evaporative and lapse rate feedbacks (e.g., \citealt{Pithan2014}).

\begin{figure}[b]
\begin{centering}
\includegraphics[scale=0.37]{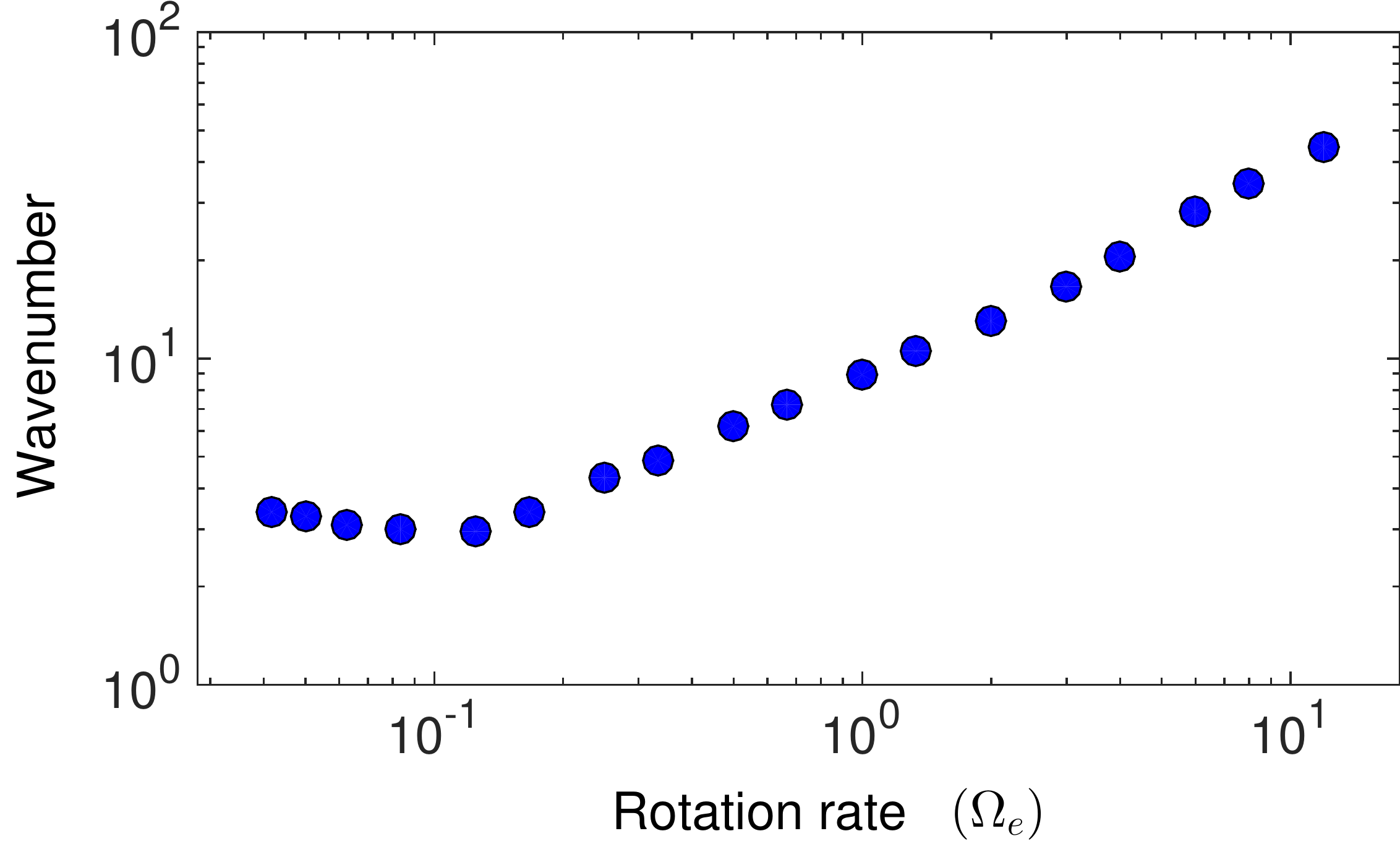}
\par\end{centering}

\caption{\label{fig:wavenumber_vs_rot} The energy-containing wavenumber as function
of the rotation rate of the planet, compared to Earth's rotation rate ($\Omega_e$). }
\end{figure}

\section{Results\label{sec:Results}}

\subsection{Dependence on the rotation rate\label{sub:Dependence-on-rotation}}

We begin with a series of experiments where we vary the rotation rate
of the planet (e.g., \citealp{Williams1982, DelGenio1987, Williams1988a,
Navarra2002, Schneider2006, Chemke2015b}). The main effect of increasing the rotation rate is
that the eddy length scales decrease (we refer to eddies as the deviation
from the time mean flow). This results from the fact that the primary
source of extratropical eddies is from baroclinic instability (e.g.,
\citealp{Pedlosky1987,Pierrehumbert1995}), and the dominant baroclinic
length scale scales inversely with rotation rate (e.g., \citealp{Schneider2006}).
This is illustrated in Fig.~\ref{fig:4 surface temp}, which shows
instantaneous snapshots of the surface temperature of experiments
with the properties of the our reference climate, but where the rotation
rate is varied between half and four times that of Earth. The waves
in the temperature field are Rossby waves (e.g., \citealp{Vallis2006}),
resulting from the latitudinal dependence of the Coriolis forces in
the horizontal momentum equations (Eq.~\ref{eq:Du},\ref{eq:Dv}). The smaller eddies in the
rapidly rotating models, demonstrated by the energy-containing
wavenumber in Fig.~\ref{fig:wavenumber_vs_rot}, are less efficient in transporting MSE meridionally,
leading to a greater equator-to-pole temperature difference for those
cases. A similar dependence has been found in other studies as well
(e.g., \citealp{Schneider2006,Kaspi2011b,Kaspi2013b}). 

\begin{figure*}
\begin{centering}
\includegraphics[scale=0.45]{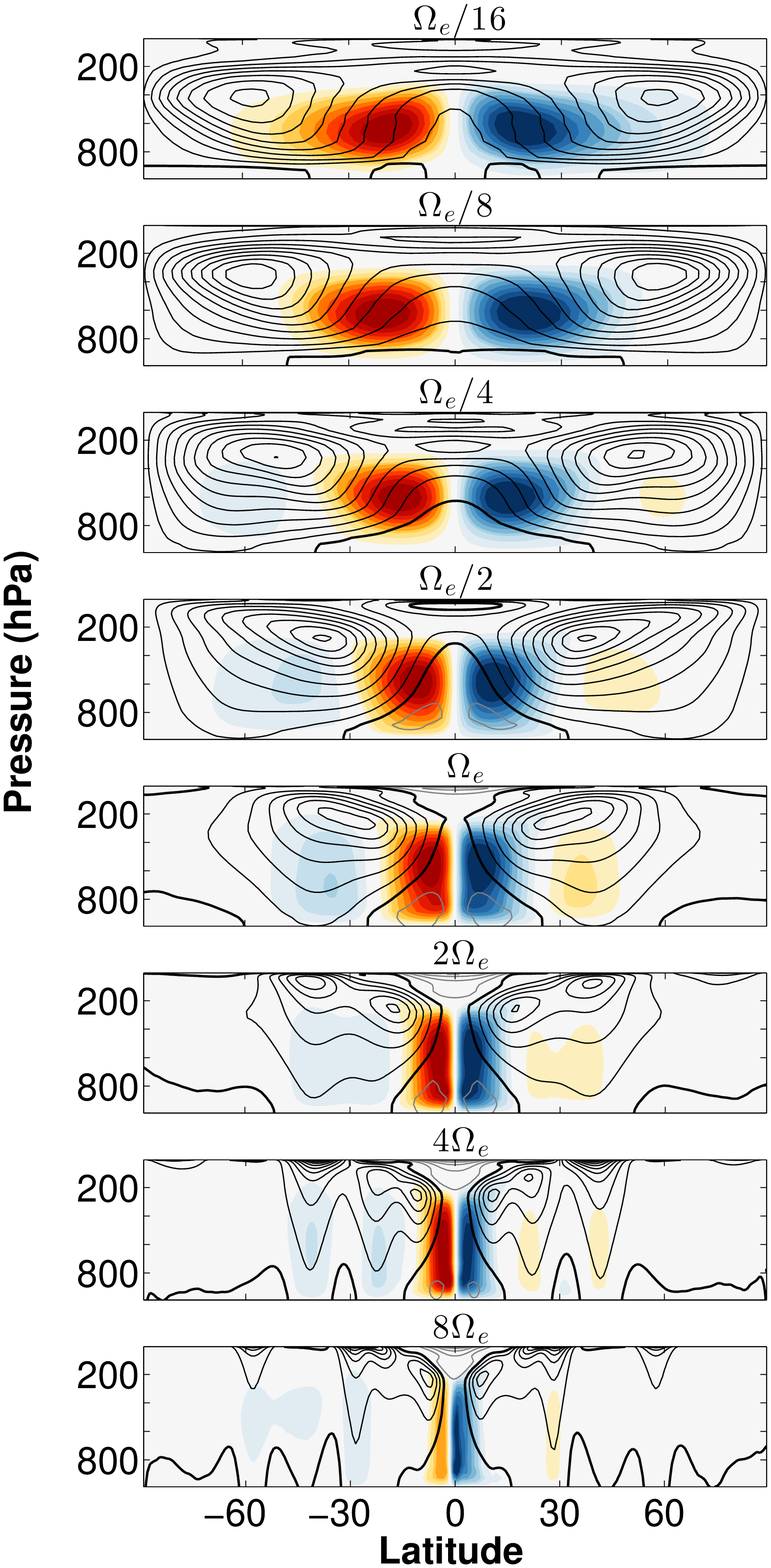} ~\includegraphics[scale=0.45]{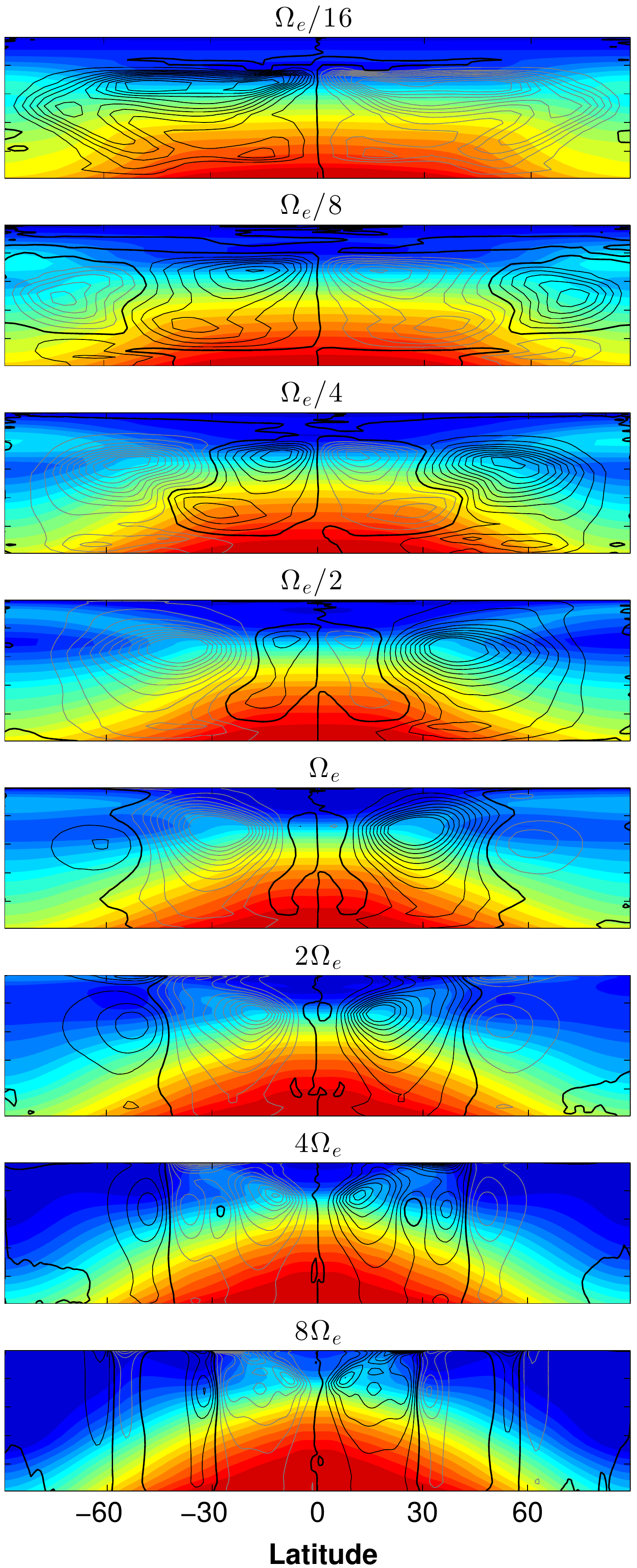}
\par\end{centering}

\caption{\label{fig:rotation_8_cases} Zonal-mean circulation for a sequence
of idealized GCM experiments ranging from 1/16th to eight times the
rotation rate of Earth from top to bottom, respectively. Left column:
Thin black contours show zonal-mean zonal wind with a contour interval
of 5~m~s$^{-1}$, and the zero-wind contour is shown in a thick
black contour. In color is the mean-meridional mass streamfunction, with
blue denoting clockwise circulation and red denoting counterclockwise
circulation. Maximum mass streamfunction values correspond to 8.7, 5.1,
4.2, 3, 1.9, 1.2, 0.6, 0.2 $\times10^{11}$~kg~s$^{-1}$,
from top to bottom, respectively. Right column: colorscale shows zonal-mean
temperature, with colorscale ranging between 210~K and 290~K. Contours
show zonal-mean meridional eddy-momentum flux, $\overline{u'v'}$.
Contour spacing grows from $9$~m$^{2}$~s$^{-2}$ in the slowly
rotating cases to $1$~m$^{2}$~s$^{-2}$ in the fast rotating cases.
Black and gray contours denote positive and negative values, respectively
(implying northward and southward transport of eastward eddy momentum,
respectively). }
\end{figure*}

The strong dependence of the atmospheric circulation 
on rotation rate is because of the dominance of the Coriolis acceleration
in the momentum balance. This can be quantified in terms of the Rossby
number which is the ratio between the typical velocity and the rotation
rate times a typical length scale $\frac{u}{\Omega L}$ \citep{Pedlosky1987}.
Typically, away from the equator for Earth's atmosphere and ocean
the Rossby number is smaller than one, meaning that in Eq.~(\ref{eq:Du}--\ref{eq:Dw}),
the leading order balance is geostrophic, and thus between the Coriolis
and pressure forces in the horizontal momentum equations. However, planetary atmospheres are not necessarily
in that regime (e.g., \citealp{Showman2014}). Solar system examples
are Venus and Titan which have rotation rates of 243 days and 16 days
respectively, and therefore are characterized by larger Rossby numbers.
This is demonstrated in Fig.~\ref{fig:rotation_8_cases}, where we
show the zonal winds, temperature, meridional mass streamfunction, and
meridional zonal-momentum fluxes ($\overline{u'v'}$) for cases of $1/16$, $1/8,$ $1/4$, $1/2,$ $1$,
2, 4 and 8 times the rotation rate of Earth. The mass streamfunction is
defined as $\psi\left(\theta,p\right) = 2\pi a\cos\theta \int v\left(\theta,p\right)dp/g$. Simulation results
presented here and throughout the paper have been both zonally
(longitudinally) and time averaged.

\begin{figure}
\begin{centering}
\includegraphics[scale=0.31]{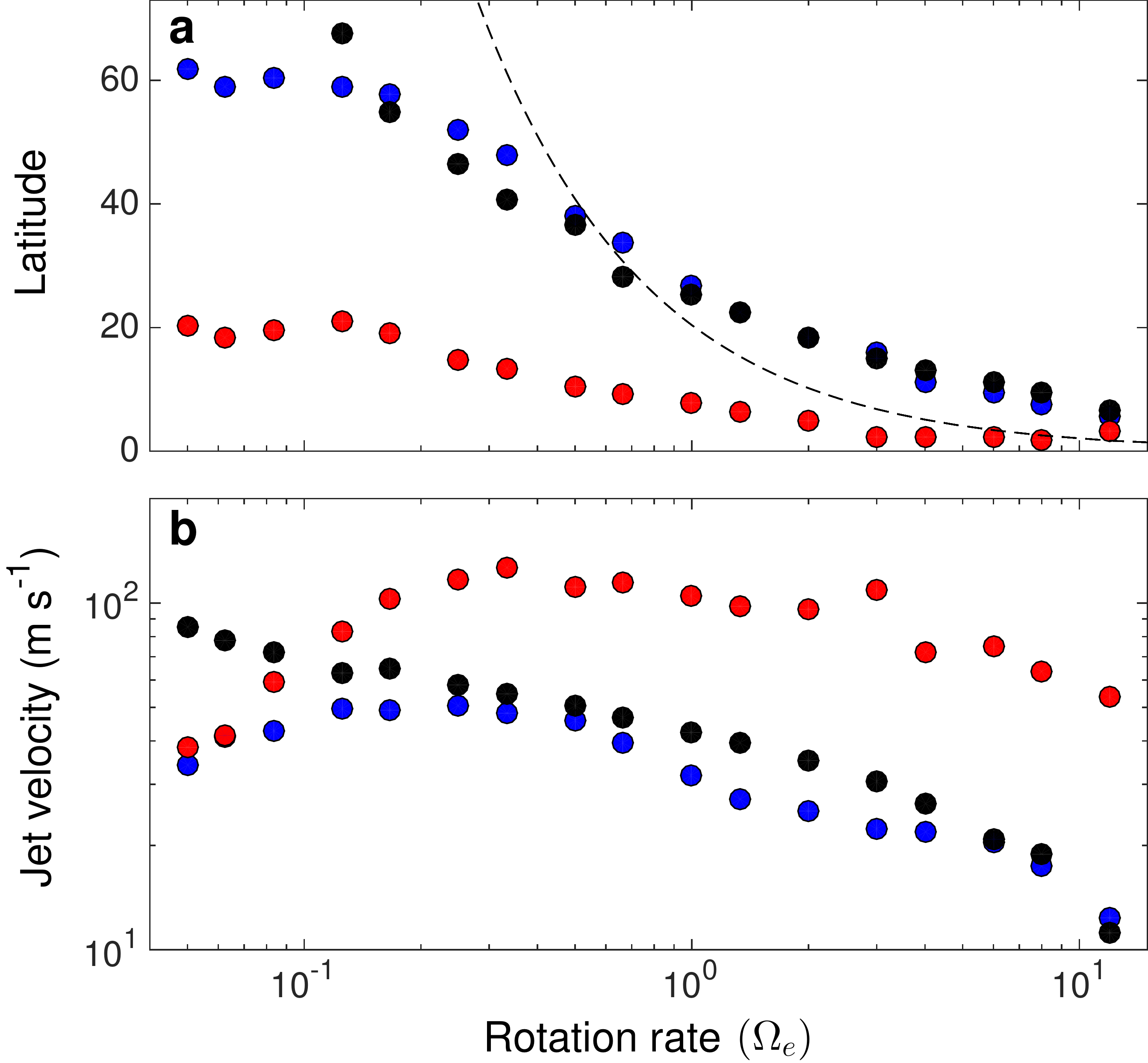}
\par\end{centering}

\caption{\label{fig:4 hadley_lat} (a) The latitude of the Hadley cell maximum
(red), Hadley cell width (black) and the latitudinal location of the
maximum jet. The dashed line shows the Hadley cell width following the
axisymmetric theory of \citet{Held1980}. (b) The magnitude of the subtropical jet (blue), the
magnitude of an angular momentum conserving wind ($u_{M}$) at the
latitude of the maximal jet (red), and the strength of the Hadley Cell
streamfunction (black) to the power of $2/5$. }
\end{figure}

\begin{figure}[b]
\begin{centering}
\includegraphics[scale=0.32]{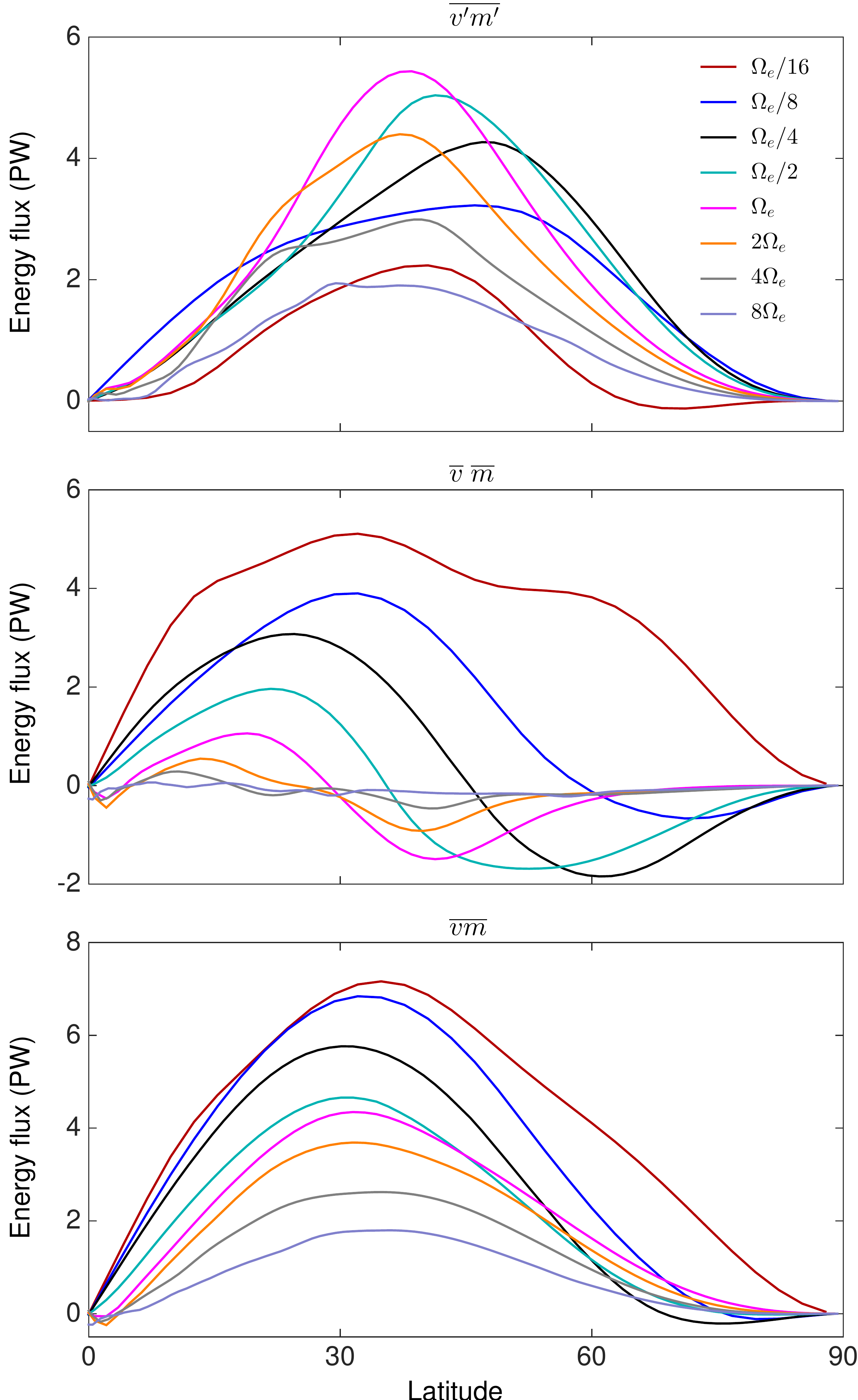}
\par\end{centering}

\caption{\label{fig:rotation_mse_flux} The eddy (top), mean (middle) and total
(bottom) poleward MSE flux as function of latitude for simulations
with different planetary rotation rates.}
\end{figure}

As rotation rate is increased the Rossby number becomes smaller and
more jets (regions of localized zonal velocity) develop (e.g.,
\citealp{Williams1982,Williams1988a,Schneider2006,Chemke2015b}).  These jets
are generated by two distinct mechanisms, which we can categorize
as thermally, and eddy, driven. In the former, the differential
heating between low and high latitudes causes
air to rise at the equator, flow 
poleward aloft, and sink at higher latitudes forming Hadley cells (colors
on the left column of Fig.~\ref{fig:rotation_8_cases}). Air flowing polewards
within the upper branch of the Hadley cell moves closer to the axis
of rotation of the planet, and therefore to conserve angular momentum
it must develop eastward velocity (contours in the left side of 
Fig.~\ref{fig:rotation_8_cases}). Within the Hadley cells meridional temperature
differences are weak, creating a strong temperature gradient on the
poleward side of the Hadley cell. For the cases of small Rossby number
(rapidly rotating) geostrophy then implies that the atmosphere must
be in thermal wind balance, meaning that 
\begin{eqnarray}
f\frac{\partial u}{\partial p} & = & \frac{R_{d}}{p}\frac{1}{a}\left(\frac{\partial T}{\partial\theta}\right)_{p},\label{eq:thermal-wind}
\end{eqnarray}
thus that the vertical wind shear is proportional to the latitudinal
temperature gradients along isobars. This then implies that the strong
temperature gradient at the poleward side of the Hadley cell must
be balanced by strong local wind shear, which then results in a local
maxima of eastward velocity (Fig.~\ref{fig:rotation_8_cases}). This
jet is referred to as the subtropical jet (due to its location on
Earth --- at the edge of the Hadley cell), and is evident in the faster
rotation cases in Fig.~\ref{fig:rotation_8_cases}. For the slower
rotation cases (Rossby number~$\gtrsim$~1), the eastward jets are
a consequence of the conservation of angular momentum of poleward
moving air in the upper branch of the Hadley cell, and therefore the
jet magnitude is closer to that expected simply by conservation of
angular momentum given by $u_{M}=\Omega a\frac{\sin^{2}\theta}{\cos\theta}$
\citep{Held1980,Vallis2006}. Figure~\ref{fig:4 hadley_lat} shows
both the magnitude of the subtropical jet and the magnitude of $u_{M}$
for a series of simulations with different rotation rates ranging
from $1/24$ to $12$ times the rotation rate of Earth. It shows that
the slowly rotating cases develop stronger jets which are close
to the angular momentum conserving value, while for the fast rotating
cases the geostrophically balanced jets (Eq.~\ref{eq:thermal-wind})
are much weaker than $u_M$. The strength of the subtropical
jet scales nicely with the Hadley cell strength to the $2/5$ power
following \citet{Held1980}, as shown in Fig.~\ref{fig:4 hadley_lat}b.

The other type of jets --- eddy driven jets --- appear in the rapidly
rotating cases. Here, breaking of Rossby waves in the extratropics
results in eddy momentum flux convergence in the extratropics\footnote{We define extratropics as regions
of the atmosphere with Rossby number $\ll 1$, (e.g., \citealp{Showman2014}).}, and
the formation of jets due to this convergence of momentum. This can
be seen in Fig.~\ref{fig:rotation_8_cases} showing multiple zonal
jets (left side) corresponding to areas where there is momentum flux
convergence (Fig.~\ref{fig:rotation_8_cases}, right side). The resulting
jets from this mechanism have a more barotropic structure then the
subtropical jets \citep{Vallis2006}. For the Earth rotation rate
case the subtropical and eddy-driven jets appear almost merged in
Fig.~\ref{fig:rotation_8_cases} (and Fig.~\ref{fig:compare_ncep_model},
but are then clearly separable for the cases with faster rotation
rates. The number of jets in each hemisphere is then related to the
typical eddy length scale, and the inverse energy cascade length scales
\citep{Rhines1975,Rhines1979,Chemke2015b}.

At slow rotation rates, the Hadley cells are nearly global, the subtropical
jets reside at high latitude, and the equator-pole temperature difference
is small (Fig.~\ref{fig:rotation_8_cases}). The low-latitude meridional momentum flux is equatorward,
leading to equatorial superrotation (eastward winds at the equator)
in the upper troposphere \citep{Mitchell2010}, qualitatively similar
to that on Venus and Titan. At faster rotation
rates, the Hadley cells and subtropical jets contract toward the equator,
and simultaneously an extratropical zone, with eddy-driven jets, develops
at high latitudes, and the equator-pole temperature difference is
large. The low-latitude meridional momentum flux is poleward, resulting
from the absorption of equatorward-propagating Rossby waves coming
from the extratropics. It is also evident from Fig.~\ref{fig:rotation_8_cases}
that as the rotation rate is increased the equator to pole temperature
difference increases. This is a result of the combination of the facts
that as rotation rate increases, the decreases in eddy length scale
results in less eddy transport polewards, and that the slower
rotation rate cases have large planetary scale Hadley cells which increase
the heat transport by the mean meridional circulation. 

\begin{figure}[b]
\begin{centering}
\includegraphics[scale=0.37]{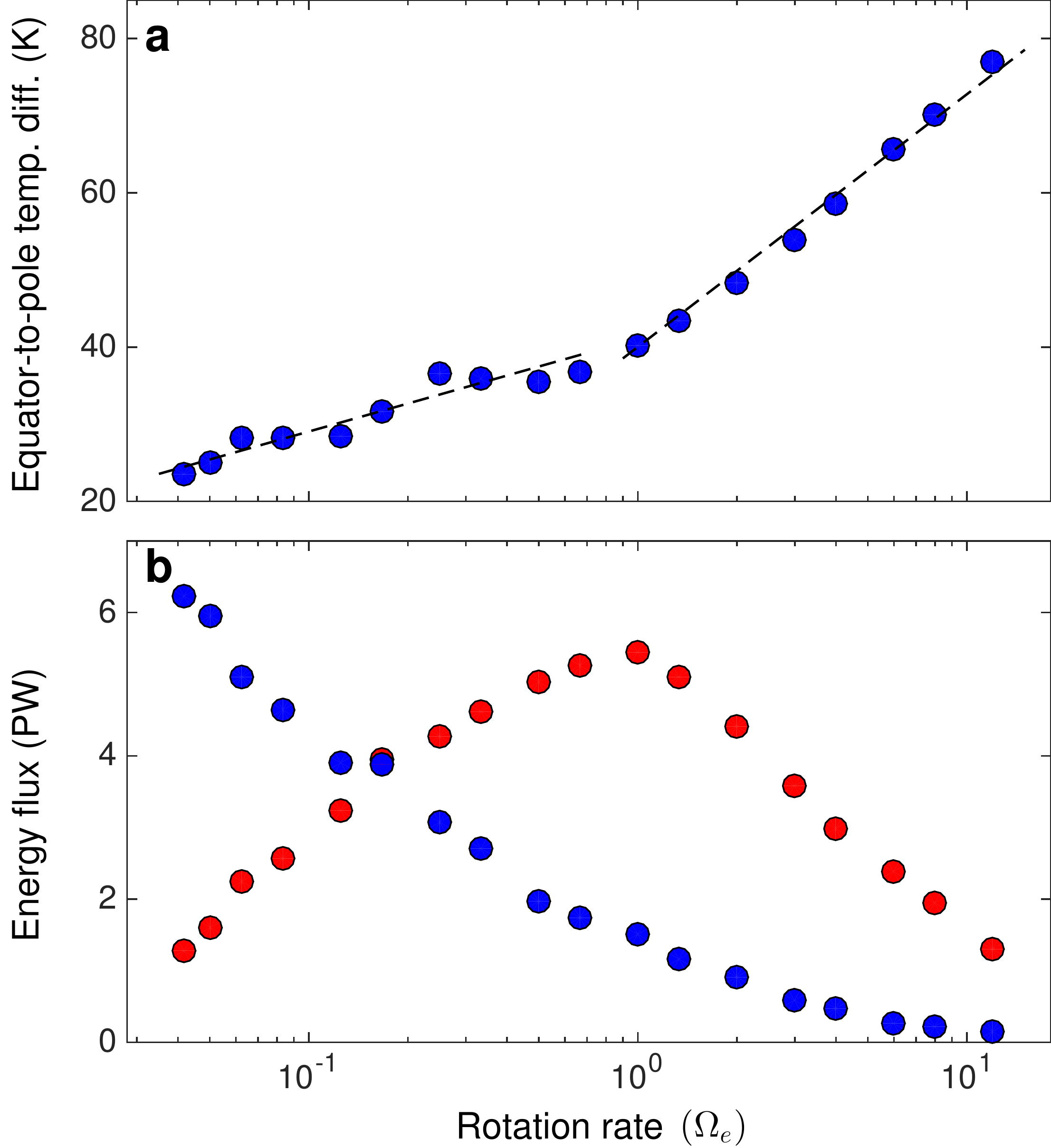}
\par\end{centering}

\caption{\label{fig:delT_rotation} (a) The equator-to-pole
surface temperature difference as function of the rotation rate of
the planet. (b) The maximum value of the poleward eddy
heat transport (red), $\overline{v'm'}$, and mean transport (blue),
$\overline{v}\,\overline{m}$, as function of the rotation rate of the
planet. Fast rotation rate simulations are dominated by the decrease
in eddy length scale with rotation rate resulting in less eddy heat
transport poleward. Slower rotation rate experiments are dominated
by large Hadley cells resulting in large mean heat transport and therefore
smaller equator-to-pole temperature differences.}
\end{figure}

These two effects are shown in Fig.~\ref{fig:rotation_mse_flux},
which shows $\overline{v}\,\overline{m}$, $\overline{v'm'}$ and
$\overline{vm}$ as function of latitude for different rotation rate
cases. It shows that the turbulent heat flux $\overline{v'm'}$ has
a nonmonotonic response to change of rotation rate. For slow rotation
rates there is weak baroclinic instability and therefore the atmosphere
is less turbulent resulting in weaker eddy heat transport ($\overline{v'm'}$),
while for fast rotation rate the atmosphere is strongly baroclinically
unstable but baroclinic zones are narrow and eddies are small resulting
in weak $\overline{v'm'}$. Thus the maximum eddy poleward heat transport
in these experiments is for rotation rates roughly similar to Earth's
(of course we should remember that our reference climate is tuned
to Earth's observed dynamics). On the other hand, the mean fluxes
decrease monotonically with the increase of rotation rate (Fig.~\ref{fig:rotation_mse_flux}b),
and this is because the Hadley cells become smaller for faster
rotation (Fig.~\ref{fig:4 hadley_lat}a). The majority of the contribution
of the mean fluxes is at low latitudes (within the Hadley cell), and
becomes a global heat transport only for the very slowly rotating
cases where the Hadley cell becomes of the order of the size of the
planet. In fact, for rotation rate cases faster than $1/4$~$\Omega_{e}$
the heat transport in the extratropics is negative because of the
existence of Ferrel cells, and it dominates the overall mean transport
($\overline{v}\,\overline{m}$). Only for the extremely slow rotation
planets when the Hadley cell is global does the extratropical heat
transport by the mean become positive
(Fig.~\ref{fig:rotation_mse_flux}b). An axisymmetric
picture predicts that the Hadley cell width should be proportional
to $\Omega^{-1}$ \citep{Held1980}, however this is complicated by
the existence of eddies leading to more complex theories regarding
where exactly the Hadley cells terminate (e.g.,
\citealp{Schneider2006a,Caballero2007, Levine2011}).
The clear dependence of the Hadley cell width on the rotation rate
is shown in Fig.~\ref{fig:4 hadley_lat}a, with the $\Omega^{-1}$
line as a reference. In addition, warming of the extratropical surface
temperature leads to additional radiative heating of the
atmosphere creating therefore a positive feedback
\citep{Pithan2014}. In the steady state shown here all these processes
have reached equilibrium.  

\begin{figure}
\begin{centering}
\includegraphics[scale=0.32]{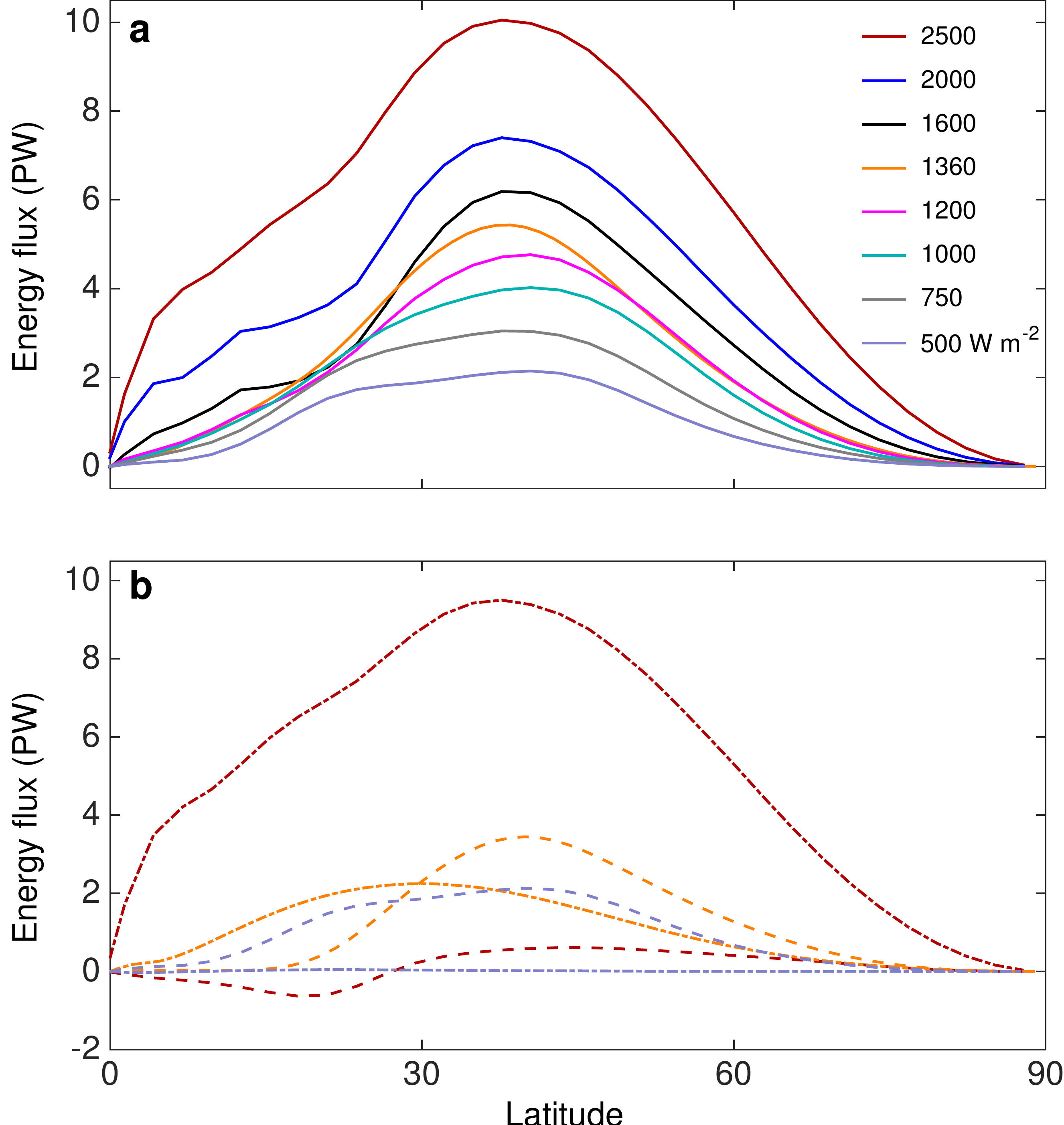}
\par\end{centering}

\caption{\label{fig:solar_mse_flux} (a) The eddy MSE transport
for experiments with different stellar flux ranging from 500 to
2500~W~m$^{-2}$ (b) Three cases from above (solar constant of 500,
1360 and 2500~W~m$^{-2}$) showing the latent heat transport (dash-dot)
and dry static energy transport (dash). This shows the strong
nonlinear effect of water vapor on the MSE transport.}
\end{figure}

\begin{figure*}
\begin{centering}
\includegraphics[scale=0.5]{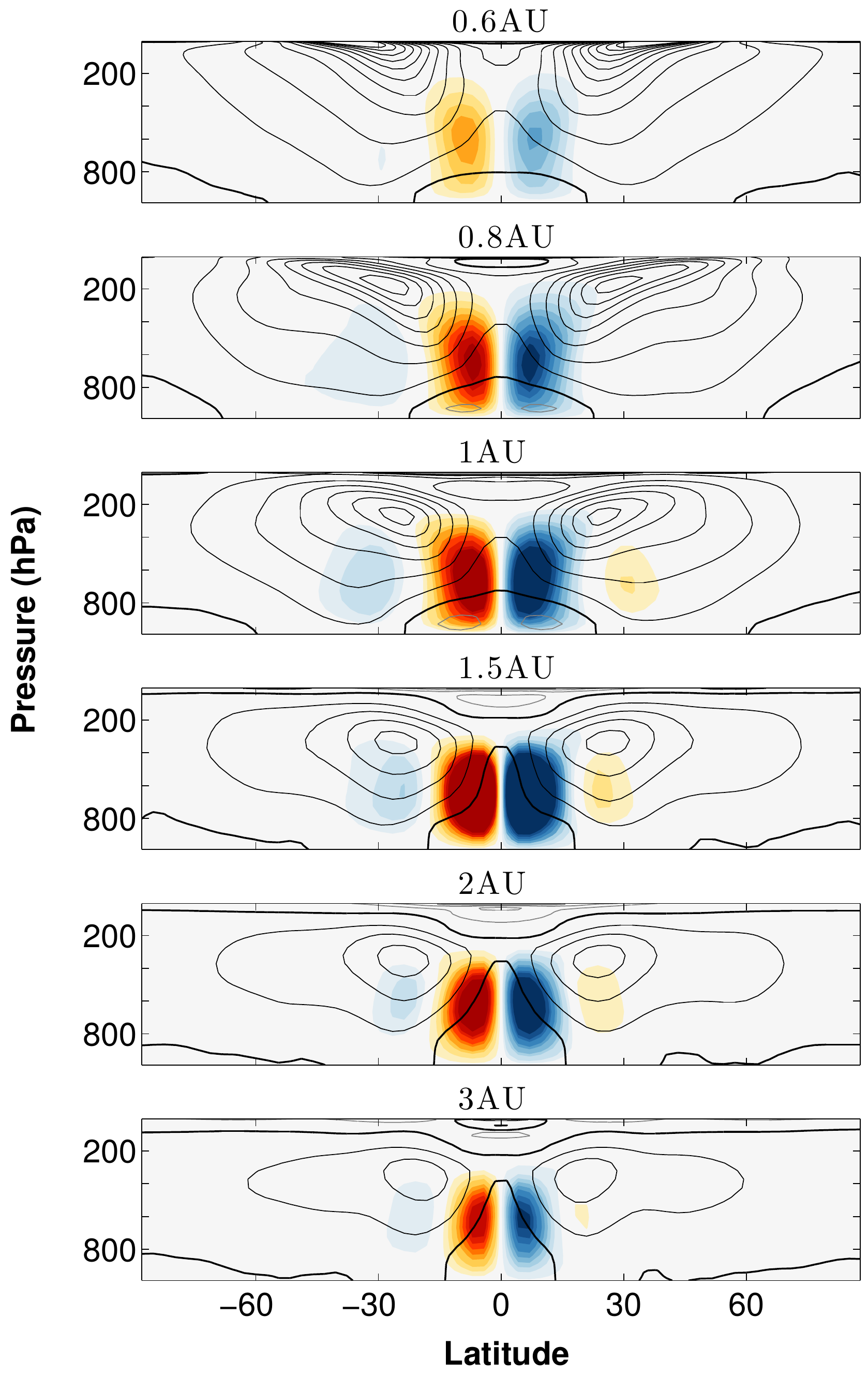}~\includegraphics[scale=0.5]{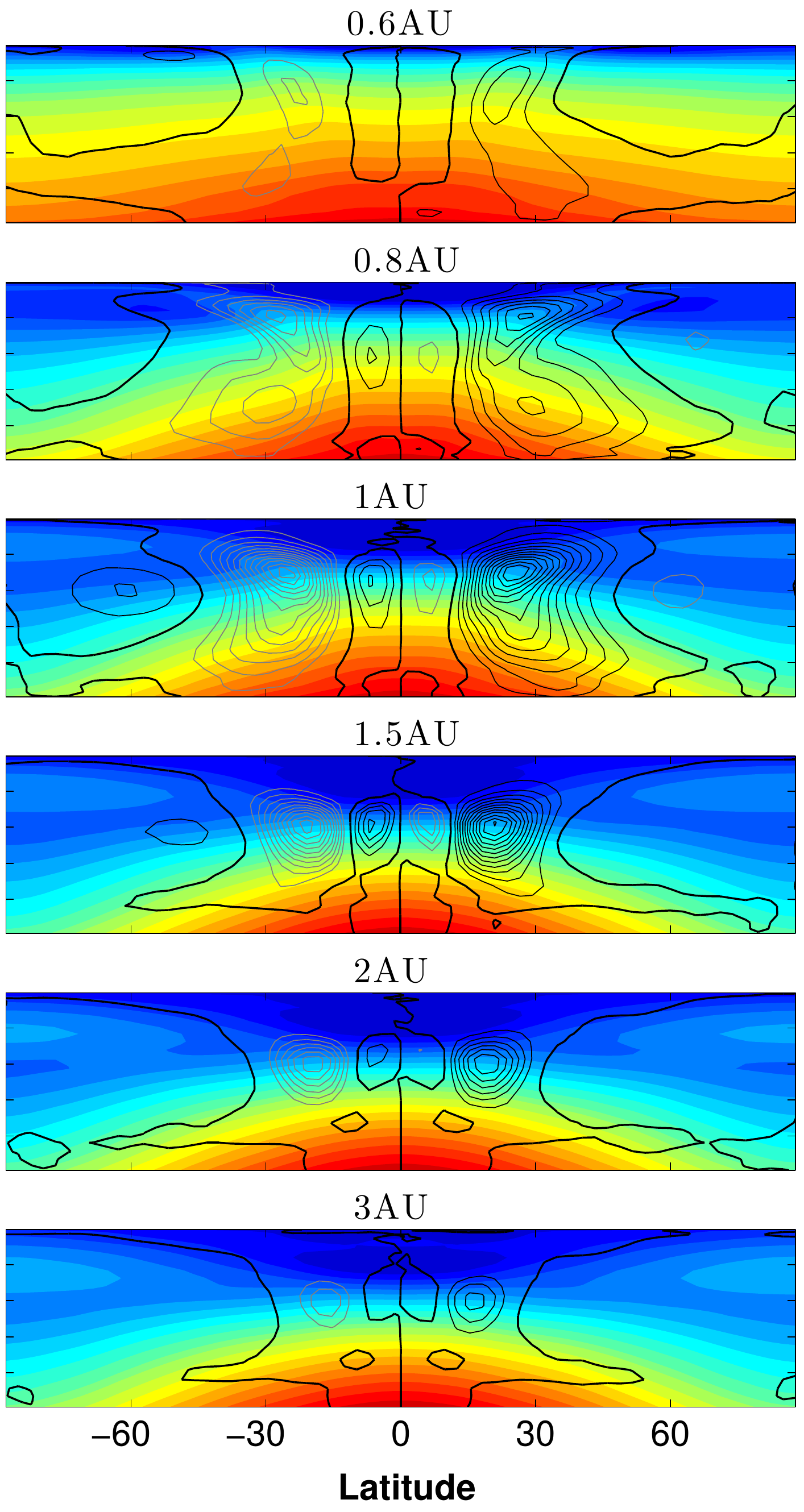}
\par\end{centering}

\caption{\label{fig:solar_5_cases} Zonal-mean circulation for a sequence of
idealized GCM experiments with stellar flux values of 3800, 2100, 1360,
607, 342 and 152 W~m$^{-2}$, which correspond to a distance of 0.6,
0.8, 1, 1.5, 2 and 3~AU from their parent star (for Solar like luminosity)
from top to bottom, respectively. Left column: Thin black contours
show zonal-mean zonal wind with a contour interval of 5~m~s$^{-1}$,
and the zero-wind contour is shown in a thick black contour. In color
is the mean-meridional mass streamfunction, with blue denoting clockwise
circulation and red denoting counterclockwise circulation. Maximum
and minimum streamfunction values correspond to $\pm2.2\times10^{11}$~kg~s$^{-1}$
respectively, for all panels. Right column: colorscale shows zonal-mean
temperature. Due to the large range of temperatures colorscale is
different for each of the panels: maximum values are 354, 318, 301,
257, 222 and 183~K, and minimum values are 270, 233, 211, 173, 159
and 124~K from top to bottom, respectively. Contours show zonal-mean
meridional eddy-momentum flux, $\overline{u'v'}$. Contour spacing
is $3$~m$^{2}$~s$^{-2}$. Black and gray contours denote positive
and negative values, respectively (implying northward and southward
transport of eastward eddy momentum, respectively). }
\end{figure*}

The sum of the eddy and the mean transports, giving the total heat
transport, results in an overall larger heat transport for slower
rotating planets; and therefore, planets with faster rotation to have
larger equator-to-pole temperature differences (Fig.~\ref{fig:delT_rotation}a).
For the fast rotating cases this total is dominated by the eddies (even
for the extreme fast rotation cases because the mean transport in those
cases is negligible), and dominated by the mean transport for the less
turbulent, slowly rotating cases.  The transition between eddy- and mean-flow
dominated transport occurs at a rotation rate of $0.2\Omega_e$ (location
where red and blue points cross in Fig.~\ref{fig:delT_rotation}b).
Interestingly, however, the dependence of equator-pole temperature difference
on rotation rate (i.e., the slope of the line in Fig.~\ref{fig:delT_rotation}a) exhibits a clear kink 
at larger rotation rates of $\Omega_e$.  This occurs because the dependence
of eddy energy flux on rotation rate changes sign at $\Omega_e$.  At 
rotation rates smaller than $\Omega_e$, eddies transport more energy when
rotation rate is larger, but at rotation rates larger than $\Omega_e$, they
transport less energy when rotation rate is larger (Fig.~\ref{fig:delT_rotation}a, red points).
By comparison, the mean flow transports less energy when rotation rate is
larger across the entire parameter range explored (Fig.~\ref{fig:delT_rotation}b, blue points).
The sum of the red and blue points implies that the {\it total} energy transport
is rather insensitive to rotation rate when $\Omega < \Omega_e$ but depends
strongly on rotation when $\Omega > \Omega_e$.  In turn, this leads to
a relatively flat dependence of equator-pole temperature difference on rotation
at low rotation rate but a strong dependence at high rotation rate, explaining
the kink in Fig.~\ref{fig:delT_rotation}a.

\begin{figure}
\begin{centering}
\includegraphics[scale=0.36]{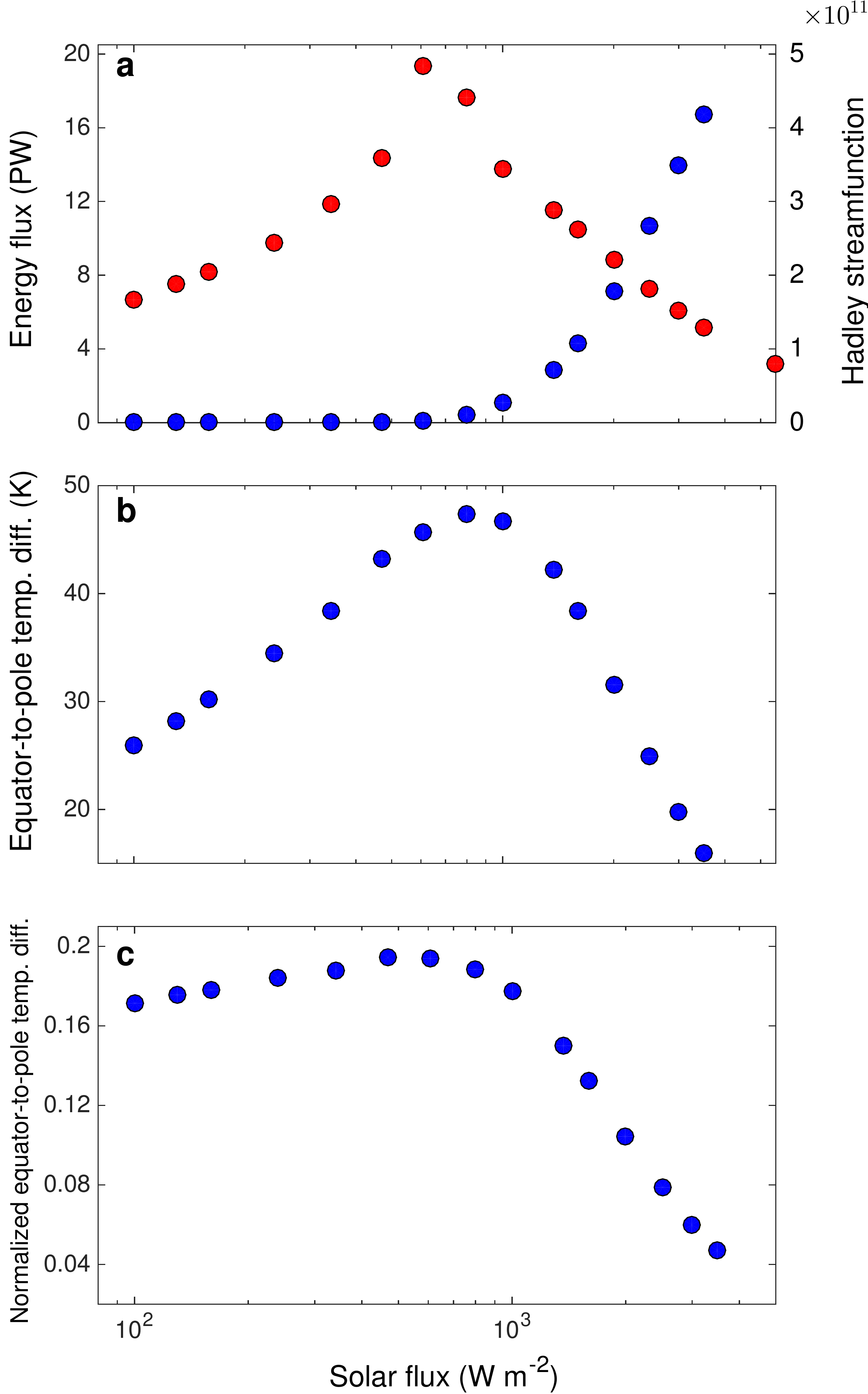}
\par\end{centering}

\caption{\label{fig:delT_solar} (a) The maximum value of the Hadley
 cell mass streamfunction
(red) and eddy latent heat flux (blue), $L\overline{v'q'}$,  as function
of the stellar heat flux. (b) The equator-to-pole surface temperature
difference, and (c) the equator-to-pole surface temperature
difference normalized by the mean surface temperature as function of the stellar heat flux. Strongly irradiated
planets which are warmer and therefore have moister atmospheres, have
a larger equator-to-pole MSE transport, which reduces the equator-to-pole
temperature difference.}
\end{figure}

\begin{figure}[b]
\begin{centering}
\includegraphics[scale=0.36]{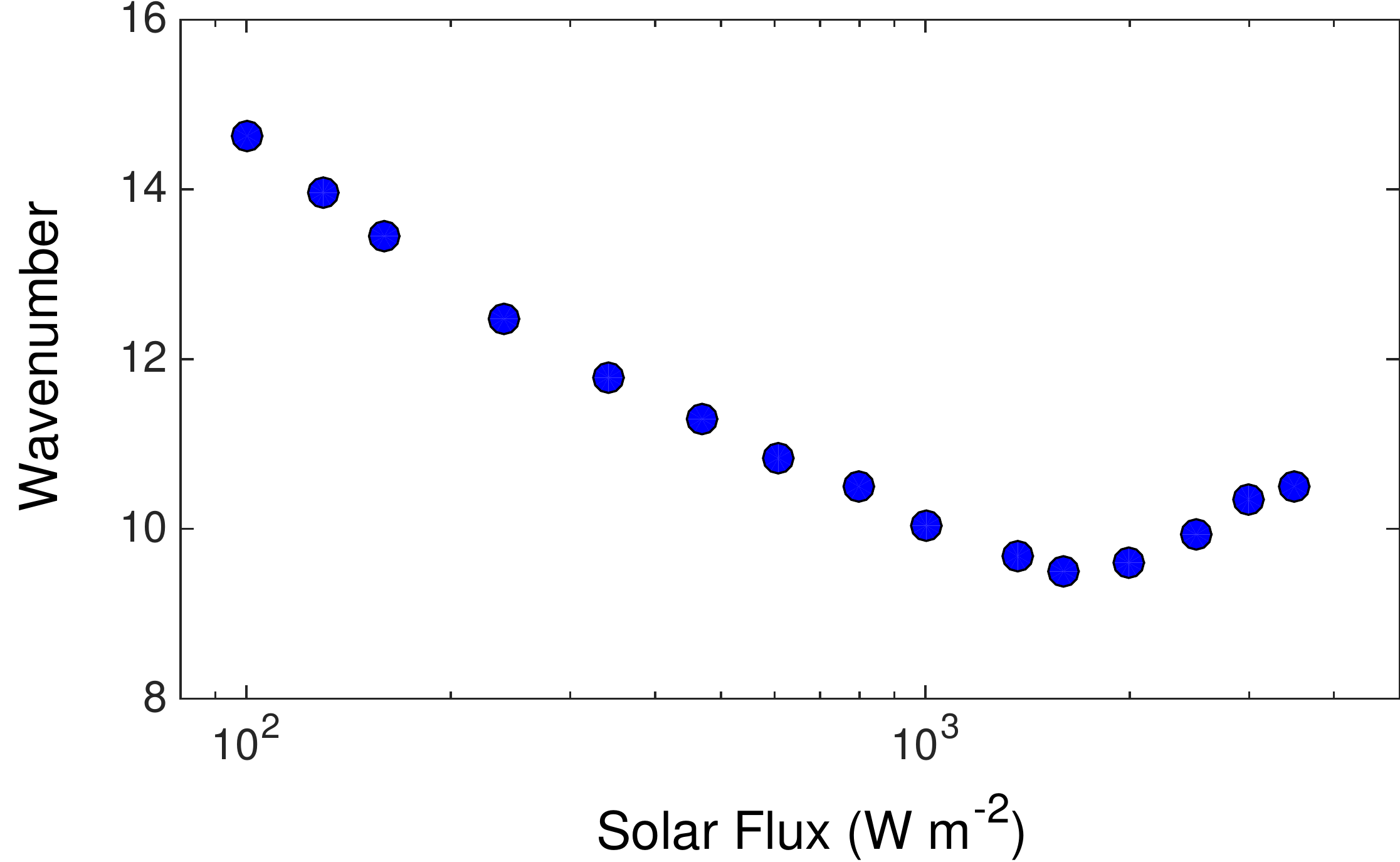}
\par\end{centering}

\caption{\label{fig:solar_wn} The energy-containing wavenumber as function of the stellar
flux.}
\end{figure}

\subsection{Dependence on stellar flux\label{sub:Dependence-on-distance}}

Next we look at the dependence of the circulation on the stellar flux.
Since these experiments have no eccentricity and we treat the parent
star as a constant point source of energy, these experiments are
equivalent to varying the distance to the parent star. In our analysis
we refer to both parameters interchangeably. As in the reference case which resembles Earth, the poleward heat
transport for all cases with different solar heat fluxes is dominated
by the eddy transport. Fig.~\ref{fig:solar_mse_flux}a shows that
planets with larger stellar flux have a larger poleward heat flux,
which results in reduction of the equator-to-pole temperature
difference (Fig.~\ref{fig:solar_5_cases}, Fig.~\ref{fig:delT_solar}b). 
The main reason for this increase in $\overline{v'm'}$
is the non-linear dependence of the Clausius-Clapeyron relation on temperature
(Eq.~\ref{eq:Clausius-Clapiron}). Warmer climates have greater 
atmospheric water vapor abundances, and therefore the relative effect
of latent heating in the total MSE transport becomes more
significant. Despite the important feedback between water vapor and
optical thickness (e.g., \citealt{Merlis2010}), for simplicity we keep 
optical thickness fixed as in the reference climate.

Fig~\ref{fig:solar_mse_flux}b shows the dry
($c_{p}\overline{v'T'}+g\overline{v'z'}$) and latent
($L\overline{v'q'}$) components of the eddy MSE transport for three
cases with different solar fluxes. For the cooler
case (500~W~m$^{-2}$), the latent heat transport is very small
compared to the dry component,
while in the Earth case they are very similar in magnitude (with the
latent component more dominant in the tropics and the dry component
more dominant in the extratropics, Fig.~\ref{fig:mse_flux}). On the
other hand in the very warm case (2500~W~m$^{-2}$) the latent
component becomes much larger than the dry component
(Fig.~\ref{fig:solar_mse_flux}b). The strong non-linearity
is expressed in the fact that the difference in heating between the
two sets of cases is similar (500, 1360 and 2000~W~m$^{-2}$), but
the increase in total heat flux has more than quadrupled.

The zonal mean climate for five cases ranging in distance from 0.6 to 2
AU (solar flux ranging between between 342 to 5470 W~m$^{-2}$) is
shown in Fig.~\ref{fig:solar_5_cases}.
Obviously the closer-in planets are much warmer (Fig.~\ref{fig:solar_5_cases},
right side). The meridional mass streamfunction however shows nonmonotonic
behavior (Fig.~\ref{fig:solar_5_cases}, left side). Planets far away
from their parent star have less thermal forcing,
and therefore Hadley and Ferrel cells get weaker as the stellar flux
decreases. However, also as the planet becomes significantly warmer
the strength of the Hadley and Ferrel cells becomes weaker. This is
due to the nonlinear dependence of water vapor on temperature, where
the increase in latent heating enables stronger MSE transport with a 
weaker circulation, resulting in weaker circulation
cells in the warmer climates. The nonlinear increase in water vapor
fluxes is also evident in Fig.~\ref{fig:delT_solar}a. For Earth like planets, the peak
in the strength of the Hadley and Ferrel cells appears at about 1.5 AU
(Fig.~\ref{fig:solar_5_cases}, left side). Quantitatively these
results depend on the temperature dependence of the
saturation vapor pressure, which can be different for planets with
different atmospheric masses since then the ratio of latent to
sensible heat will be different. However, despite the fact that the
turning point may be different (Fig.~\ref{fig:delT_solar}a), we expect
the general behavior to be similar to the results presented here.

\begin{figure*}
\begin{centering}
\includegraphics[scale=0.5]{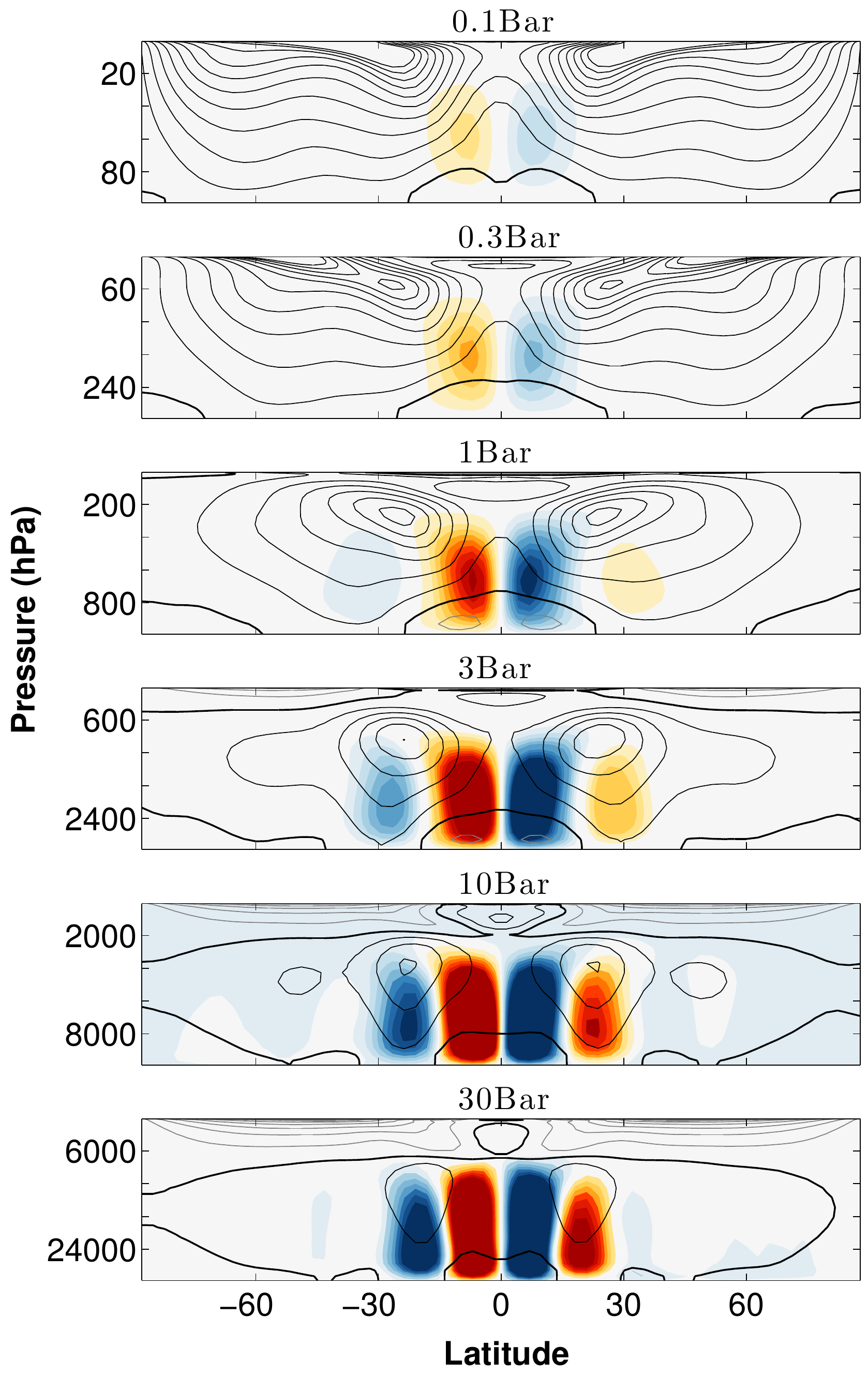} ~\includegraphics[scale=0.5]{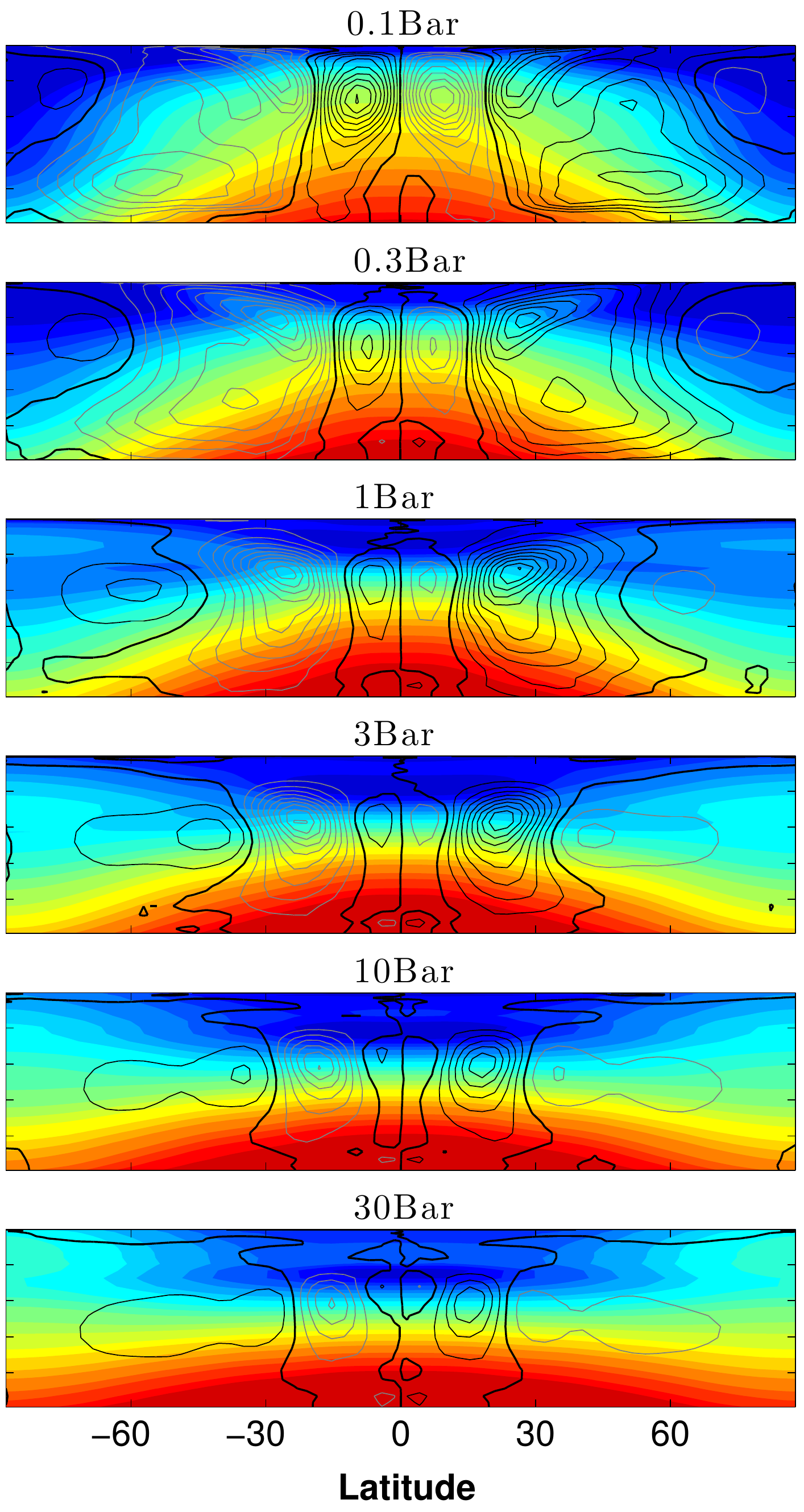}
\par\end{centering}

\caption{\label{fig:ps_5_cases} Zonal-mean circulation for a sequence of idealized
GCM experiments with surface pressure ranging from 0.1 to
30 bars from top to bottom, respectively. Left column: Thin black
contours show zonal-mean zonal wind with a contour interval of 5~m~s$^{-1}$.
The zero-wind contour is shown in a thick black contour. In color
is the mean-meridional mass streamfunction, with blue denoting clockwise
circulation and red denoting counterclockwise circulation. Maximum
and minimum streamfunction values correspond to $\pm3\times10^{11}$~kg~s$^{-1}$,
respectively for all panels. Right column: colorscale shows zonal-mean
temperature, with colorscale ranging between 210~K and 290~K. Contours
show zonal-mean meridional eddy-momentum flux, $\overline{u'v'}$.
Contour spacing grows from $6$~m$^{2}$~s$^{-2}$ in the less massive
cases to $1$~m$^{2}$~s$^{-2}$ in the more massive cases. Black
and gray contours denote positive and negative values, respectively
(implying northward and southward transport of eastward eddy momentum,
respectively). }
\end{figure*}

The domination of the moist component of the MSE flux results in that warmer
planets have a smaller equator-to-pole temperature differences as
long as moisture plays an important role in the transport as in the
warm example in Fig.~\ref{fig:solar_mse_flux}b. However, we find that
once moisture is less important (cooler climates), this relation reverses
and then as the distance to the star is increased the equator-to-pole
temperature difference decreases again (Fig.~\ref{fig:delT_solar}b). Thus, despite
the decrease in MSE flux with deceasing stellar flux
(Fig.~\ref{fig:solar_mse_flux}a), the equator-to-pole difference decreases leading to 
the nonmonotonic dependence in Fig.~\ref{fig:delT_solar}b. Mostly, this is due to the fact that for
the cooler planets the mean temperature is smaller, and therefore even
if the relative temperature difference would not have changed the
absolute temperature difference between equator and pole is
smaller. However, even the normalized equator-to-pole temperature
difference (normalizing by the average surface temperature) shows a
small decrease of temperature difference with reduction of solar 
flux (Fig.~\ref{fig:delT_solar}c), which
is due to the increase in radiative time constant for the colder planets.
Note that for these simulations the energy-containing
wavenumber generally decreases with solar flux 
(Fig.~\ref{fig:solar_wn}). Therefore unlike the rotation 
rate experiments (section~\ref{sub:Dependence-on-rotation})
where the eddy length scale decreased with rotation
rate (Fig.~\ref{fig:wavenumber_vs_rot}), limiting the eddies ability
to transport heat poleward and causing an increase in equator-to-pole
temperature difference with rotation rate, here the
reduction of eddy length scale does not result in increased
equator-to-pole temperature difference. Thus, for the cooler planets,
despite the decrease of eddy length scale with larger distance to the
parent star, the equator-to-pole temperature difference is reduced.

\begin{figure}[b]
\begin{centering}
\includegraphics[scale=0.31]{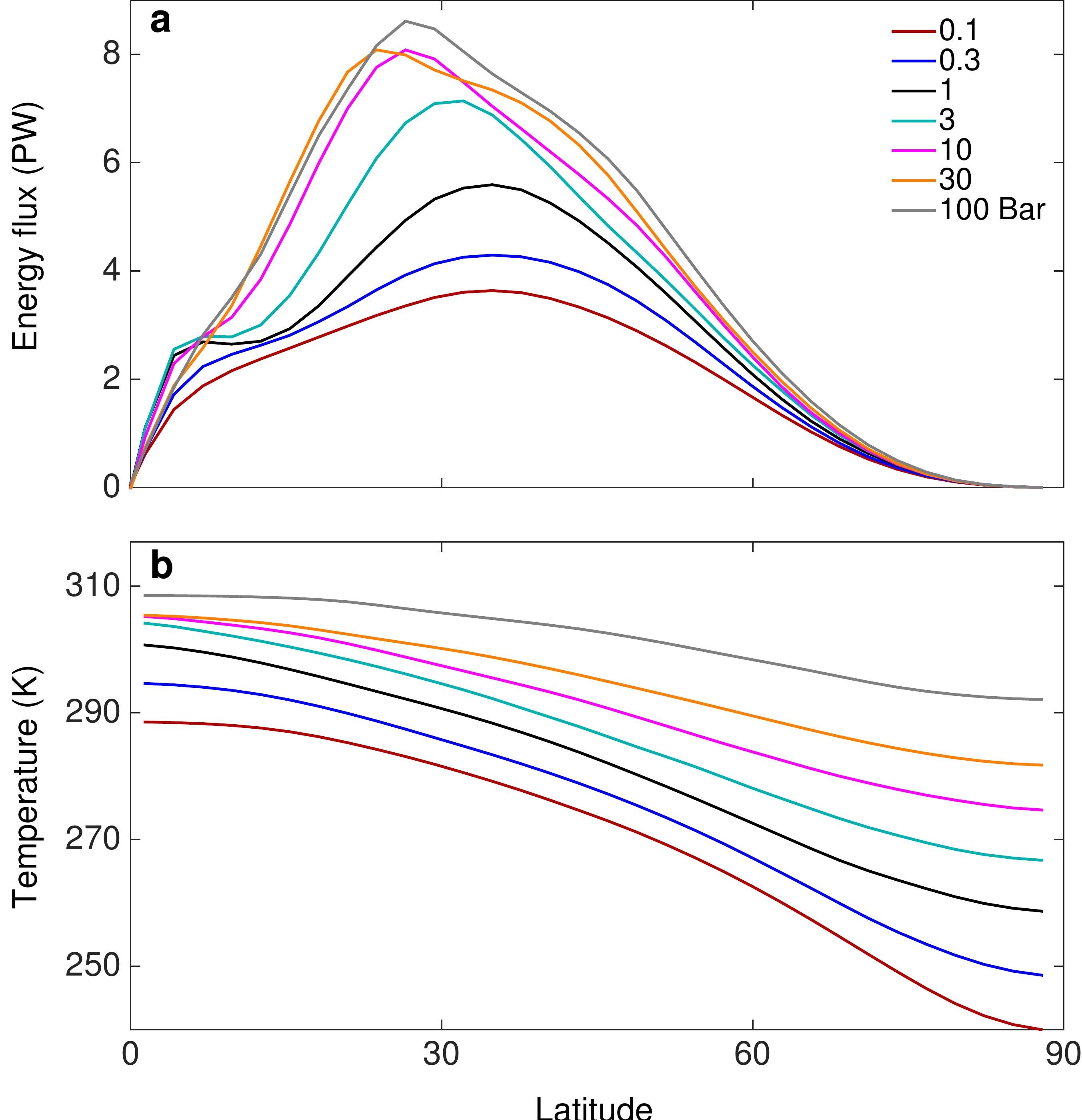}
\par\end{centering}

\caption{\label{fig:ps_mse_flux} (a) The poleward eddy MSE flux, and (b) the
surface temperature as function of latitude for simulations with different
surface pressure ranging from 0.1~bar to 100~bar. More massive atmospheres
generally have a larger poleward MSE flux resulting in a reduced equator-to-pole
temperature difference.}
\end{figure}

\begin{figure}
\begin{centering}
\includegraphics[scale=0.35]{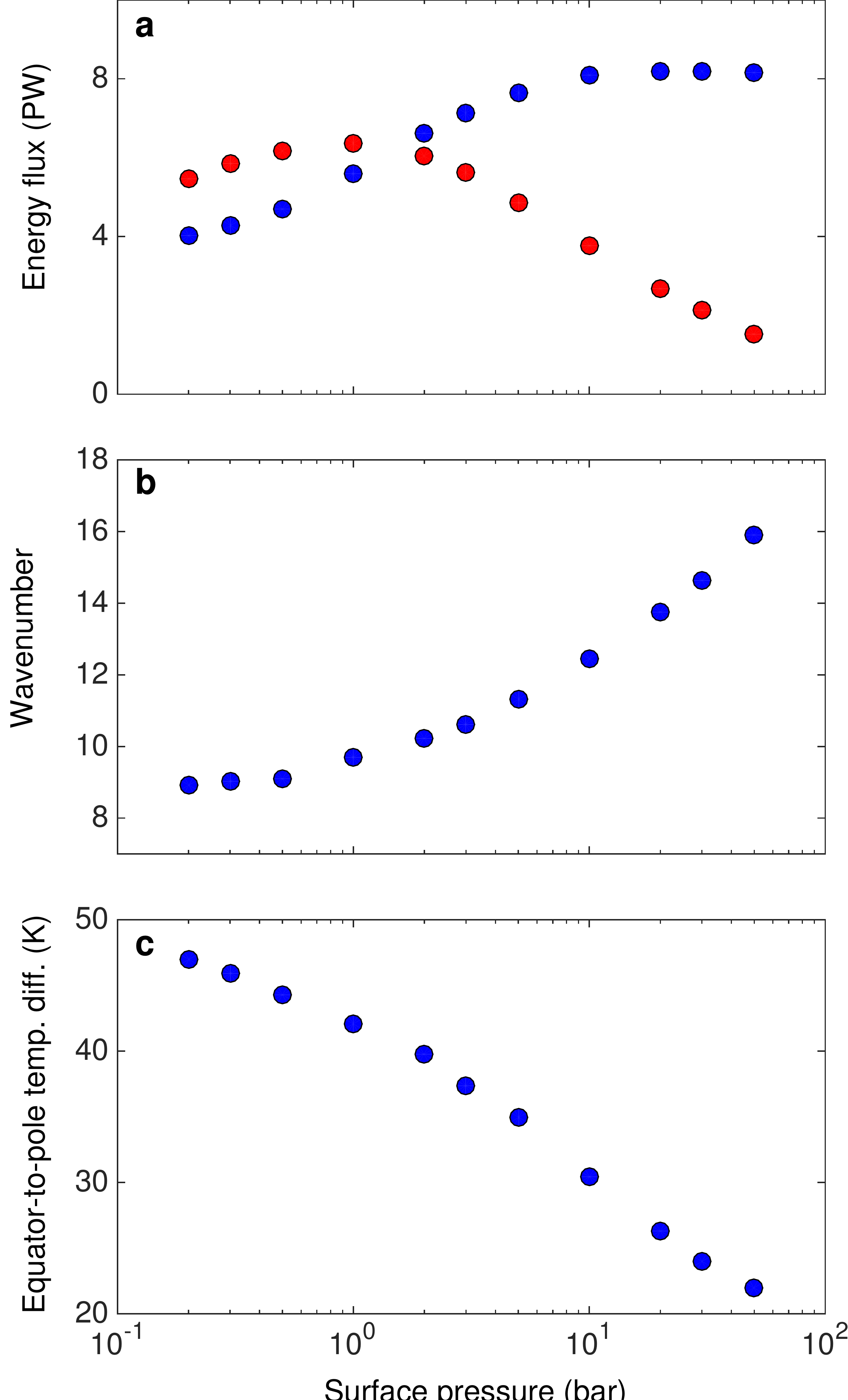}
\par\end{centering}

\caption{\label{fig:delT_ps} (a) The maximum value of the meridional
  eddy MSE flux (blue); and the total vertical MSE flux (red) averaged between 400 and 600~hPa, as function of
surface pressure. (b) The mean energy-containing wave number as function
of surface pressure. (c) The equator-to-pole surface temperature difference
as function of surface pressure. More massive atmospheres generally
have a larger poleward MSE transport resulting in a reduced equator-to-pole
temperature difference. The two competing effects of more 
MSE transport due to more atmospheric mass, and less MSE transport
because of smaller and weaker eddies level out the equator-to-pole MSE transport
in very massive atmospheres.}
\end{figure}

\subsection{Dependence on atmospheric mass\label{sub:Dependence-on-atmospheric-mass}}

Atmospheric masses can vary considerably based on the planetary composition
and history. Even in our close neighbors in the solar system atmospheric
masses vary by orders of magnitude from an atmosphere of $92$~bars
on Venus, to an atmosphere of $0.006$~bars on Mars. The other terrestrial
type atmosphere in the solar system, that of Titan, has an atmospheric surface
pressure similar to Earth of $1.5$~bar. In this section we experiment
with the atmospheric mass of the planet, by keeping all other parameters
constant as in our reference climate, and varying only surface pressure.
We use no convection scheme and do not vary optical thickness despite
increasing atmospheric mass, to allow an even comparison between the
simulations. Fig.~\ref{fig:ps_5_cases} shows that as the atmospheric
mass is increased the meridional cells of mass streamfunction become stronger and
narrower, resulting also in the subtropical jet being closer to the
equator. Note that the mass streamfunction in
Fig.~\ref{fig:ps_5_cases} is integrated over pressure, meaning that
for the case of these simulations the increase in streamfunction strength does not necessarily coincide
with stronger wind speeds. In fact, here as surface pressure is
increased the Hadley and Ferrel cells have weaker velocities despite
the increase in of the mass streamfunction. Eddy momentum
flux convergence is aligned with the jet location and also becomes
closer to the equator and weaker with increasing atmospheric
mass. Note that the contour interval of $\overline{u'v'}$ decreases
with atmospheric mass in Fig.~\ref{fig:ps_5_cases}. 

\begin{figure*}
\begin{centering}
\includegraphics[scale=0.55]{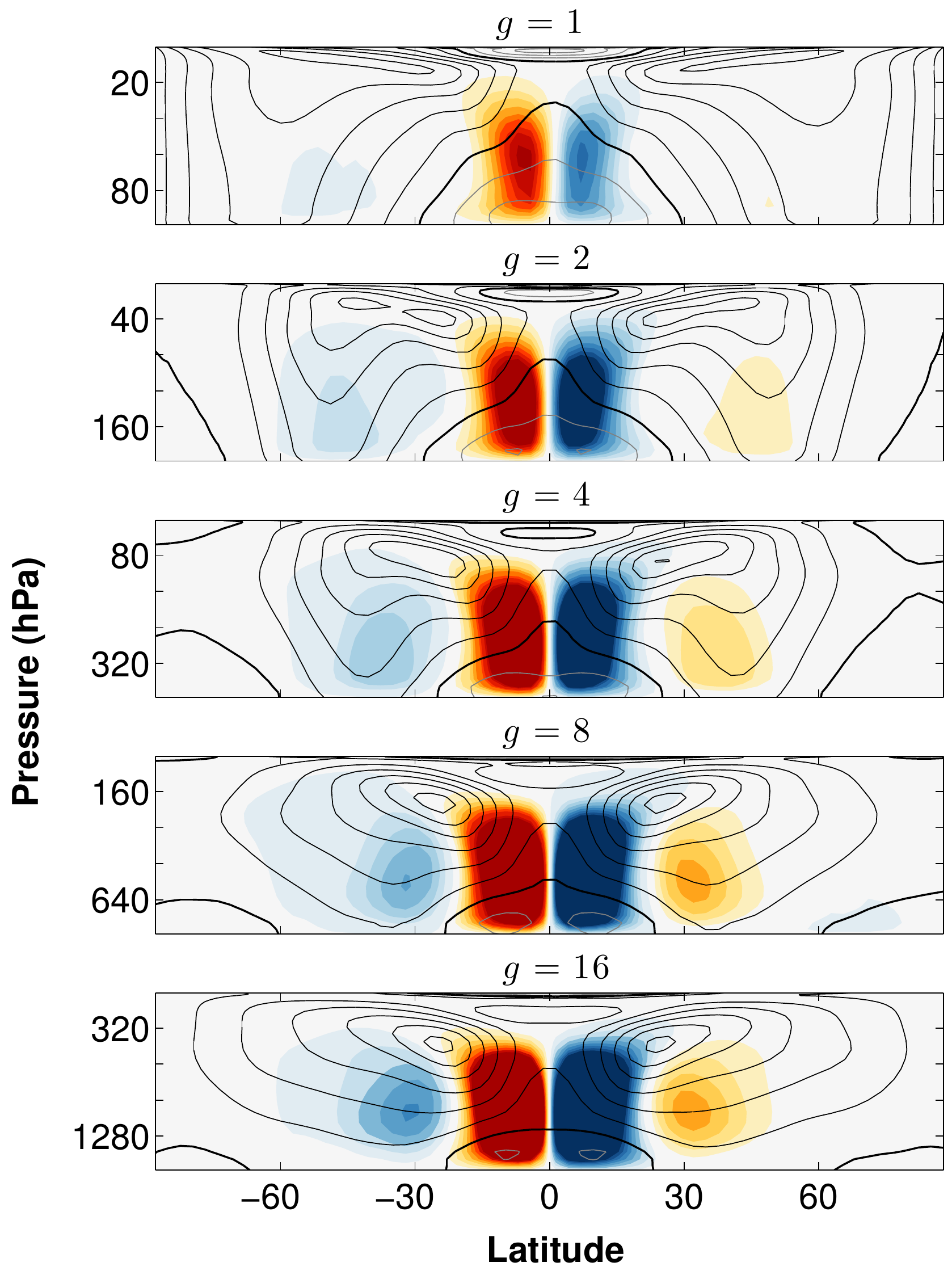}\includegraphics[scale=0.55]{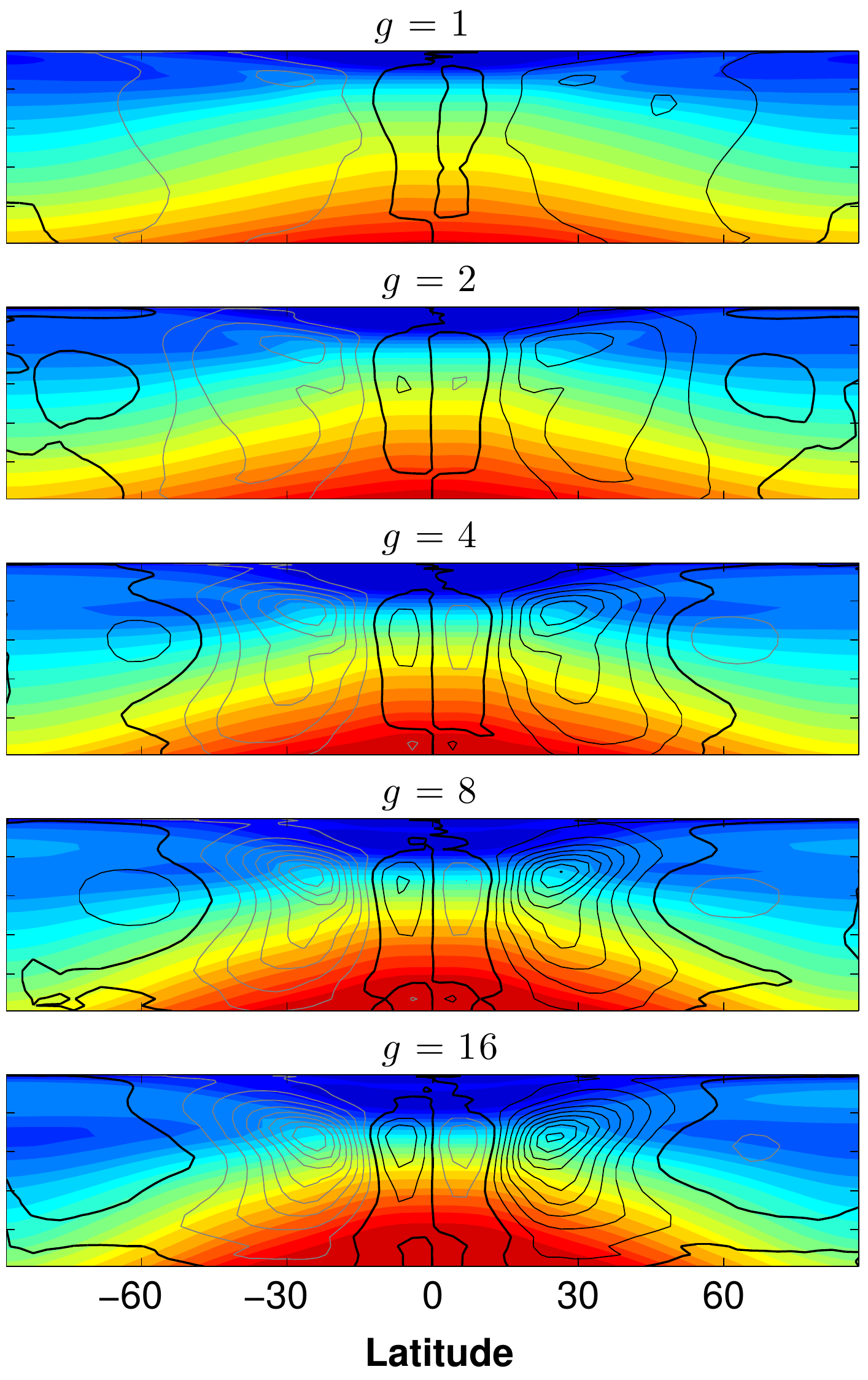}
\par\end{centering}

\caption{\label{fig:grav-5-cases}Zonal-mean circulation for a sequence of
idealized GCM experiments ranging in mean planetary density from 562~kg~m$^{-3}$
to 7,874~kg~m$^{-3}$ (corresponding to surface gravity ranging
from 1~m~s$^{-2}$ to 16~m~s$^{-2}$). Left column: Thin black
contours show zonal-mean zonal wind with a contour interval of 5~m~s$^{-1}$,
and the zero-wind contour is shown in a thick black contour. In color
is the mean-meridional mass streamfunction, with blue denoting clockwise
circulation and orange denoting counterclockwise circulation. Maximum
and minimum streamfunction values grow with the gravity, and thus
correspond to $\pm\left(1,2,4,8,16\right)\times10^{10}$~kg~s$^{-1}$,
from top to bottom respectively. Right column: colorscale shows zonal-mean
temperature, with colorscale ranging between 210~K and 290~K. Contours
show zonal-mean meridional eddy-momentum flux, $\overline{u'v'}$.
Contour spacing is $5$~m$^{2}$~s$^{-2}$. Black and gray contours
denote positive and negative values, respectively (implying northward
and southward flux of eastward eddy momentum, respectively). }
\end{figure*}

Planets with more massive atmospheres also have stronger equator-to-pole
MSE flux (Fig.~\ref{fig:ps_mse_flux}a), which reduces the equator-to-pole
temperature difference (Fig.~\ref{fig:ps_5_cases} right side), and
therefore the jet strength. In
all cases the mass flux is dominated by the eddy component (as in
the Earth case in Fig.~\ref{fig:mse_flux}), and the increase in MSE
flux follows monotonically with the increase in atmospheric mass. This
increase is mostly due to the increase in pressure (as the flux is
vertically integrated over the atmosphere), and not due to the
increase in the eddy correlations themselves ($\overline{v'm'}$) which actually
decrease with atmospheric mass.
The increase in MSE flux with atmospheric mass becomes more
gradual beyond $10$~bars, likely because of the strong decrease
in eddy length scale (Fig.~\ref{fig:delT_ps}b) for the extremely
massive atmospheres. Thus the two competing effects of more efficient
energy transport due to more atmospheric mass, and less efficient transport
because of smaller eddies level out the equator-to-pole heat transport
in very massive atmospheres. Nonetheless, the general trend of smaller
equator-to-pole temperature differences with increasing atmospheric
mass is obvious in Fig.~\ref{fig:ps_5_cases} (right column), 
\ref{fig:ps_mse_flux}b, and~\ref{fig:delT_ps}c.

Fig.~\ref{fig:ps_mse_flux}b shows that as atmospheric mass increases
surface temperatures increases as well. This is a consequence of the
fact that although horizontal heat transport increases with
increasing atmospheric mass, vertical heat transport reduces in
magnitude. This is demonstrated in Fig.~\ref{fig:delT_ps}a showing the
total vertical MSE transport in the mid atmosphere (averaged
between 400 and 600~hPa). Thus, despite the increase in the strength
of the mass streamfunction, when integrated globally more massive atmospheres have less 
vertical heat transport, which results in the surfaces
accumulating more heat (the optical thickness in these simulations
depends only on $\sigma$ and therefore is equally distributed in
height in all simulations). These differences are larger in midlatitudes
than in the tropics since at midlatitudes the eddies play a larger
role in the heat transport. This results in the fact that more massive
atmospheres have larger extratropical lapse rates, which acts to
destabilize the atmosphere. However, this increase of horizontal heat
fluxes with atmospheric mass has the opposite effect (thus stabilizing the
atmosphere), and is more dominant in our simulations, resulting in 
weaker eddy momentum fluxes and weaker jets as can be seen in 
Fig.~\ref{fig:ps_5_cases}.

\subsection{Dependence on planetary mean density\label{sub:Dependence-on-planetary-density}}

\begin{figure}
\begin{centering}
\includegraphics[scale=0.27]{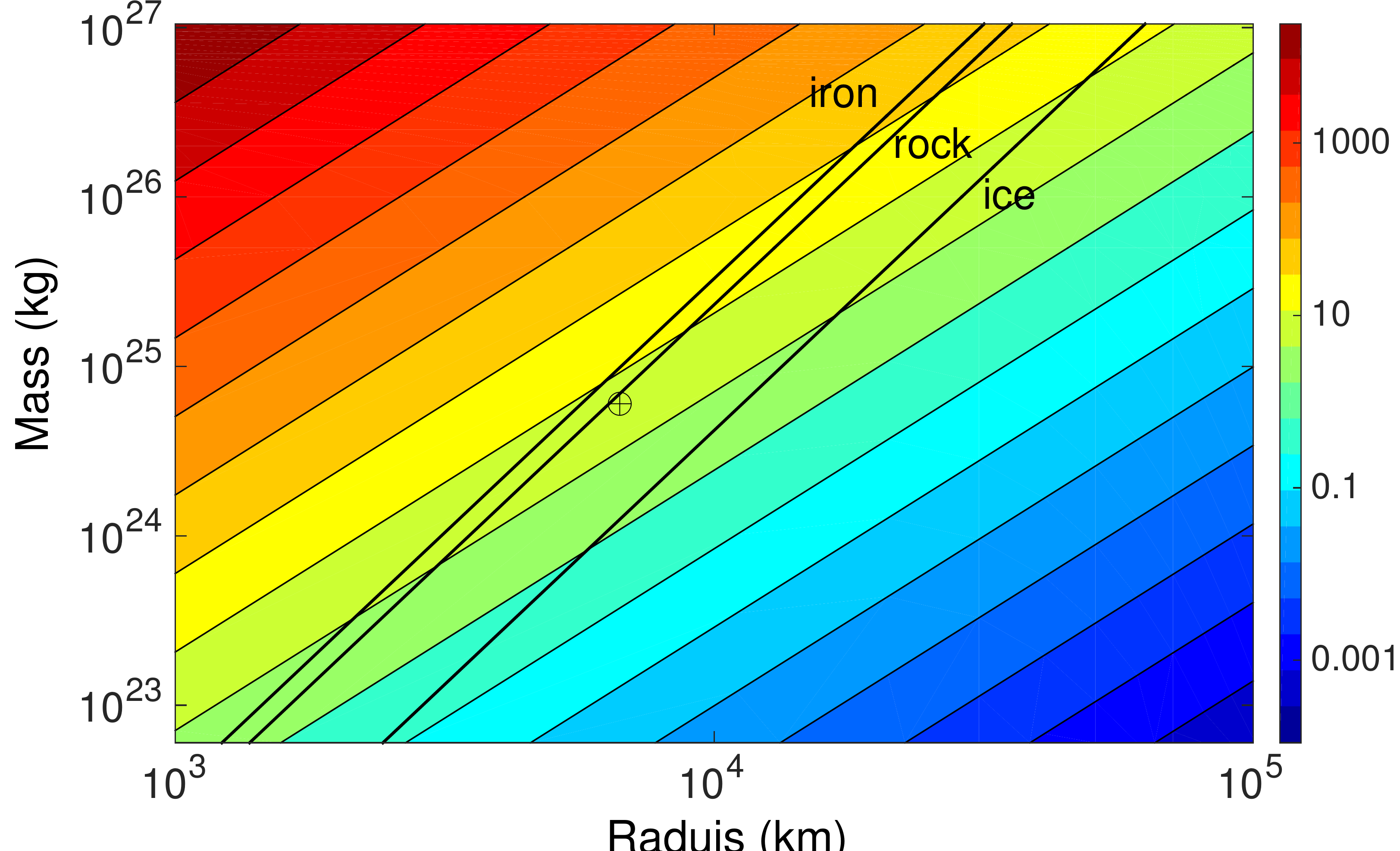}
\par\end{centering}

\caption{\label{fig:grav_RM} The surface gravity (colorscale, $\rm m\,s^{-2}$) as function of
planetary mass and radius. Black lines denote planets with a mean density of
ice (1000~kg~m$^{-3}$), rock (5520~kg~m$^{-3}$) and iron (7,874~kg~m$^{-3}$).
The $\oplus$ symbol denotes the location of Earth in this phase space. }
\end{figure}

Terrestrial type exoplanets likely span a wide range of mean densities
ranging from ice-rich planets to heavy planets composed primarily
of rock and iron.  Here we explore the effect of varying the planetary
mean density between the extremes of a relatively light ice-planet
(1000~kg~m$^{-3}$), to that of a heavy iron-planet (7,874~kg~m$^{-3}$),
while keeping the atmospheric mass constant. Thus, in these simulations
the planet radius is kept fixed to that of Earth, and we also keep
Earth's atmospheric mass by varying the surface pressure hydrostatically
given the varying surface gravity, where $g=\frac{4\pi}{3}G\rho$
(Fig.~\ref{fig:grav_RM}). Fig.~\ref{fig:grav-5-cases} shows that
Hadley and Ferrel cell strength increases with the gravity coefficient
(note that the color scale changes between the panels). For the Hadley cell case the increase in width and strength
is consistent with \citet{Held1980} suggesting a $g^{1/2}$ increase
in Hadley cell width and $g^{3/2}$ in Hadley cell intensity. In our
simulations, despite showing the same trends, the increase in Hadley
cell width and intensity is more modest likely due to the role of eddies.
Concurrently, the midlatitude jets become more subtropical in nature (thus
more baroclinic and closer to the edge of the Hadley cell). In the
water-density planet the midlatitude jets are mainly eddy driven,
weaker and more barotropic, while in the denser planets due to the
dominance of the Hadley cell, and despite the strengthening of the
eddies, the jets peak near the edge of the Hadley cells. On the other
hand, the increase in Ferrel-cell strength is likely due to stronger baroclinic
instability in midlatitudes (see below). 

\begin{figure}[b]
\begin{centering}
\includegraphics[scale=0.45]{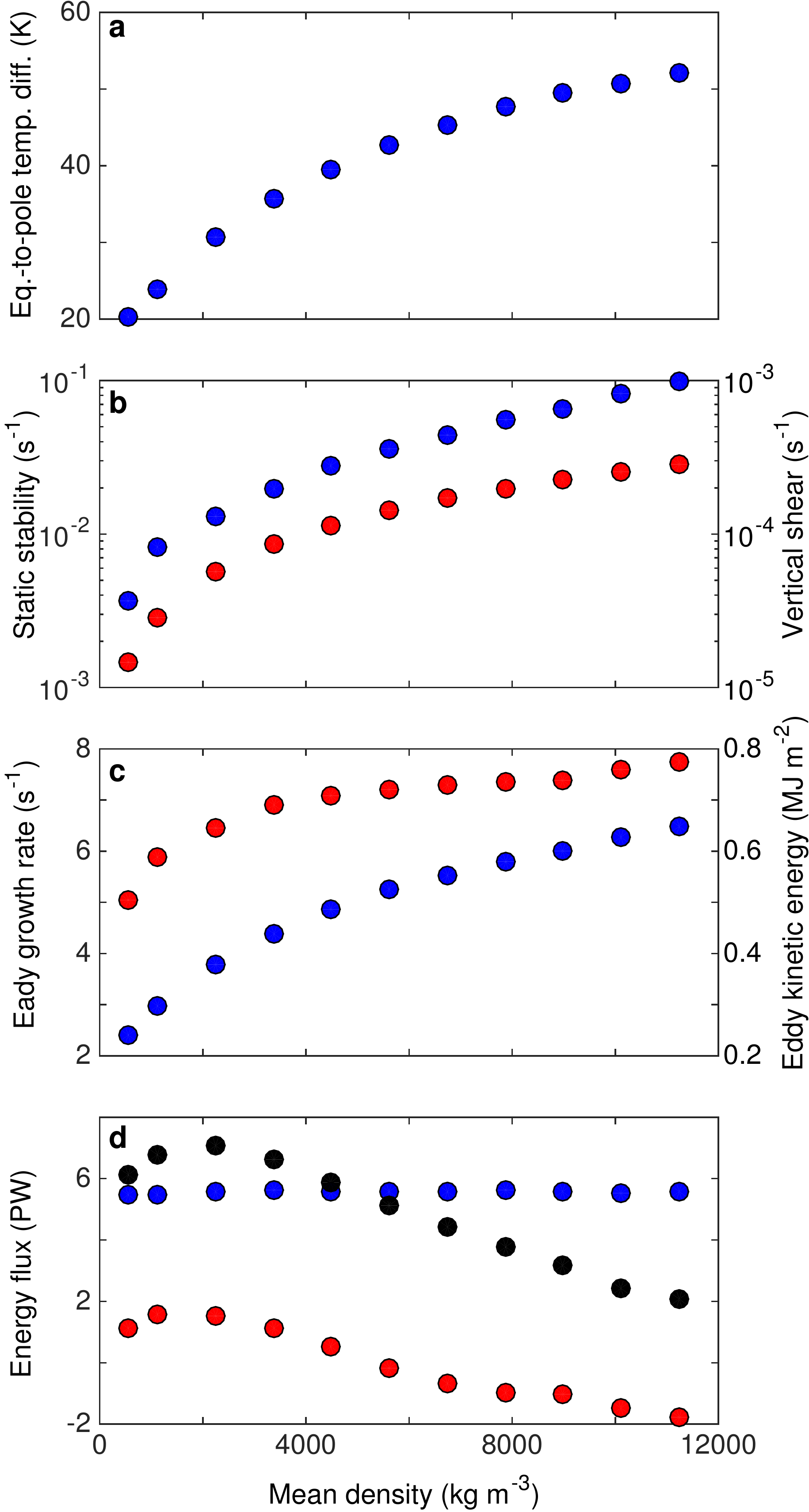}
\par\end{centering}

\caption{\label{fig:delT_grav} (a) The equator-to-pole surface temperature
difference as function of the mean density of the planet ranging between
density of an ice-planet to that of a heavy iron-planet (Fig.~\ref{fig:grav_RM}).
(b) Static stability (red), and vertical wind shear (blue) as function
of the mean density of the planet. (c) Eady growth rate (Eq.~\ref{eq: eady},
blue), and eddy kinetic energy (red). (d) Eddy MSE flux (blue), mean MSE
flux (red) and total MSE flux (black) as function of the mean planetary
density.}
\end{figure}

Unlike the rotation and atmospheric-mass experiments
(sections~\ref{sub:Dependence-on-rotation},\ref{sub:Dependence-on-atmospheric-mass}), here the increased
Hadley cell strength also results in an increase in equator-to-pole
temperature differences (Fig.~\ref{fig:grav-5-cases}, right side),
which shows roughly a tripling of the equator-to-pole surface temperature
between the ice-density planet and the iron-density planet (Fig.~\ref{fig:delT_grav}a).
The reason for this variation is that as the mean density increases,
so do the buoyancy frequency and vertical shear which affect the Eady
growth rate 
\begin{equation}
\sigma\sim\frac{\Omega\sin\theta}{N}\frac{\partial u}{\partial z},\label{eq: eady}
\end{equation}
representing the growth rate of baroclinic eddies in the atmosphere
\citep{Eady1949,Lindzen1980a}. The static stability and vertical
shear exert opposite effects on the Eady growth rate and both depend
on gravity. The static stability,
$N^{2}=\frac{g}{\Theta}\frac{\partial\Theta}{\partial z}$,  has a
dependence on gravity both because of the direct dependence on gravity and
the fact that the vertical gradient of potential temperature becomes
larger for larger mass (again, because of the reduction of the
vertical eddy heat flux). The vertical shear has a linear dependence since for a similar forcing
by a temperature gradient thermal wind implies that the vertical shear
will grow linearly with gravity (Eq.~\ref{eq:thermal-wind}).
Taking into account both dependencies (Fig.~\ref{fig:delT_grav}b), we find that heavier planets will have a
larger Eady growth rate implying stronger baroclinic eddies and eddy
kinetic energy (Fig.~\ref{fig:delT_grav}c). However, this does not
cause stronger eddy heat fluxes (Fig.~\ref{fig:delT_grav}d, blue
points), but rather the larger eddy activity in midlatitudes strengthens
the eddy-driven Ferrel cell that drives a stronger equatorward mean
circulation (Fig.~\ref{fig:delT_grav}d, red points) that weakens
the overall poleward heat transport (Fig.~\ref{fig:delT_grav}d, black
points). The reduced overall poleward heat transport then results
in an increased equator-to-pole temperature difference
(Fig.~\ref{fig:delT_grav}a). The heavier planets therefore have
stronger Ferrel cells and eddies in the extratropics, which is
also evident when looking at the strength of the eddy kinetic energy,
$\frac{1}{2g}\int\left(u'^{2}+v'^{2}\right)dp$, integrated over the troposphere (Fig.~\ref{fig:delT_grav}c).
On the other hand, if atmospheric surface pressure,
rather than atmospheric mass, is held constant while the gravity is
varying (not shown), then the simulations would resemble those of
varying atmospheric mass where the equator-to-pole temperature differences
decreases with increasing the gravity.

\begin{figure}[b]
\begin{centering}
\includegraphics[scale=0.36]{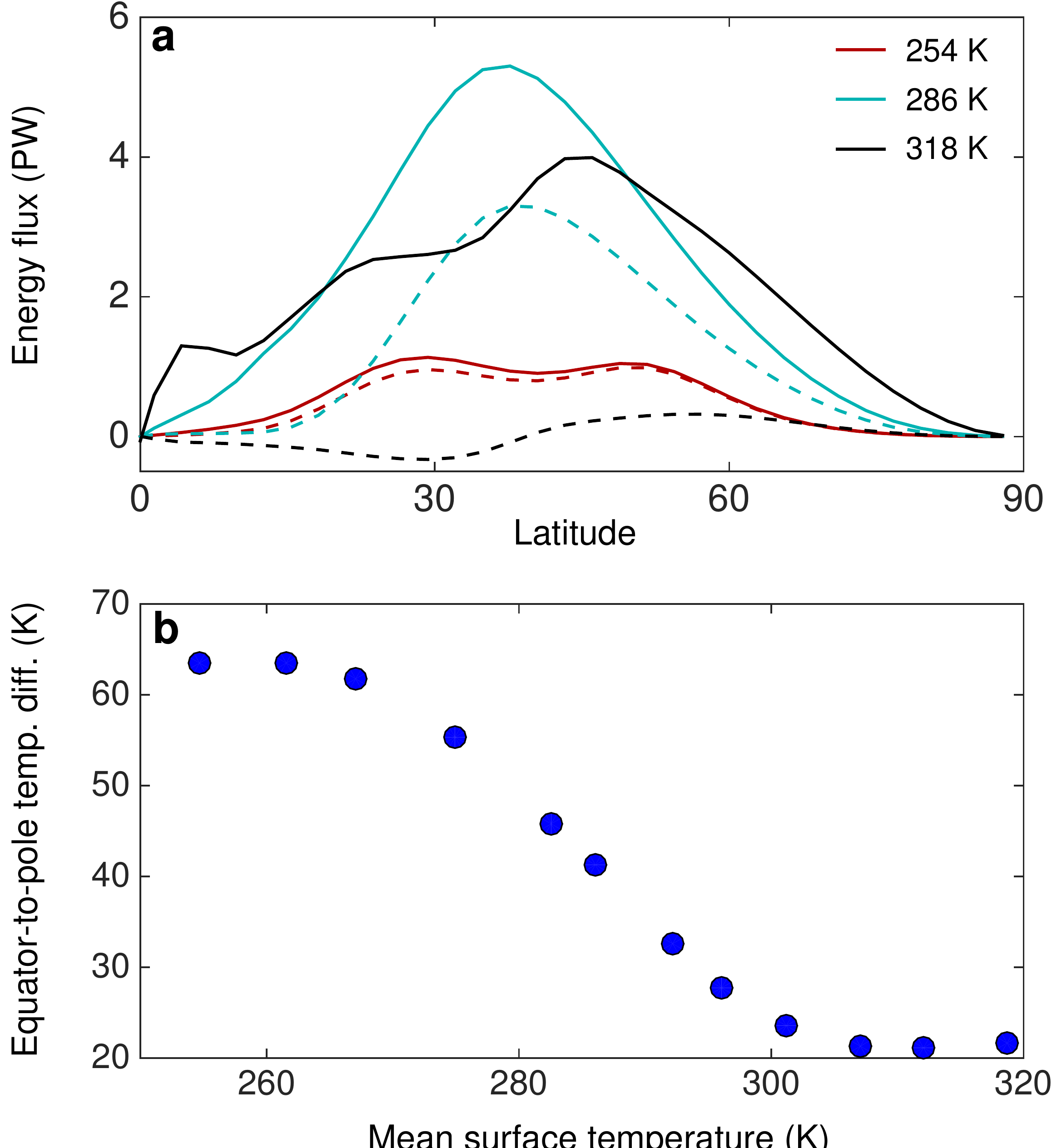}
\par\end{centering}

\caption{\label{fig:delT_tau} (a) The eddy moist (solid) and dry (dashed)
static energy fluxes as function of latitude for three cases of different
optical thicknesses, which correspond to a mean surface temperature
of 254, 286 and 318~K. (b) The equator-to-pole surface temperature
difference as function of the mean surface temperature of the planet,
corresponding to experiments with 0.1 to 15 times the optical thickness
of the reference case, which correspond to mean surface temperatures
of $253$~K to $319$~K, respectively.}
\end{figure}

\subsection{Dependence on optical thickness\label{sub:Dependence-on-optical}}

In all previous sections optical thickness (Eq.~\ref{eq:tau}) has
been kept fixed, set to the parameters that give an Earth-like climate
in the reference simulation (Fig.~\ref{fig:compare_ncep_model}).
Increasing the optical thickness while keeping other parameters fixed
(including atmospheric mass) results in an atmosphere that absorbs
more of the emitted longwave radiation and therefore is warmer. How
this effects eddy momentum transport in the atmosphere and the equator-to-pole
temperature difference is less straight forward; we explore this issue
here.  Over this series of simulations we increase $\tau_{p}$ and $\tau_{e}$
(Eq.~\ref{eq:tau}) by the same factor, thus increasing the optical
thickness linearly over all latitudes. In the series of experiments
presented here the optical thickness was varied between $0.1$ to
$15$ times the reference case, which corresponds to mean surface
temperatures of $253$~K to 319~K respectively. In resemblance to
the case of varying the stellar flux (section~\ref{sub:Dependence-on-distance}),
the results are dominated by the nonlinear response of temperature
to the atmospheric water-vapor abundance. For the cases with
low optical thickness the atmosphere is cold resulting in dry static
energy dominating the moist static energy fluxes (Fig.~\ref{fig:delT_tau}a),
and thus less overall poleward heat transport resulting in a large
equator-to-pole temperature difference (Fig.~\ref{fig:delT_tau}b).
As the optical depth is increased the moist component of the MSE becomes
larger resulting in stronger MSE transport and therefore smaller equator-to-pole
temperature differences. However, unlike the stellar flux experiments
the monotonic increase does not happen over all latitudes resulting
that at midlatitudes the colder simulation have stronger flux. This
requires further investigation. Note that in our simple model as optical
thickness does not vary with the amount of water vapor, this model
does not allow for a runaway effect as perhaps relevant to Venus. 

\begin{figure}
\begin{centering}
\includegraphics[scale=0.34]{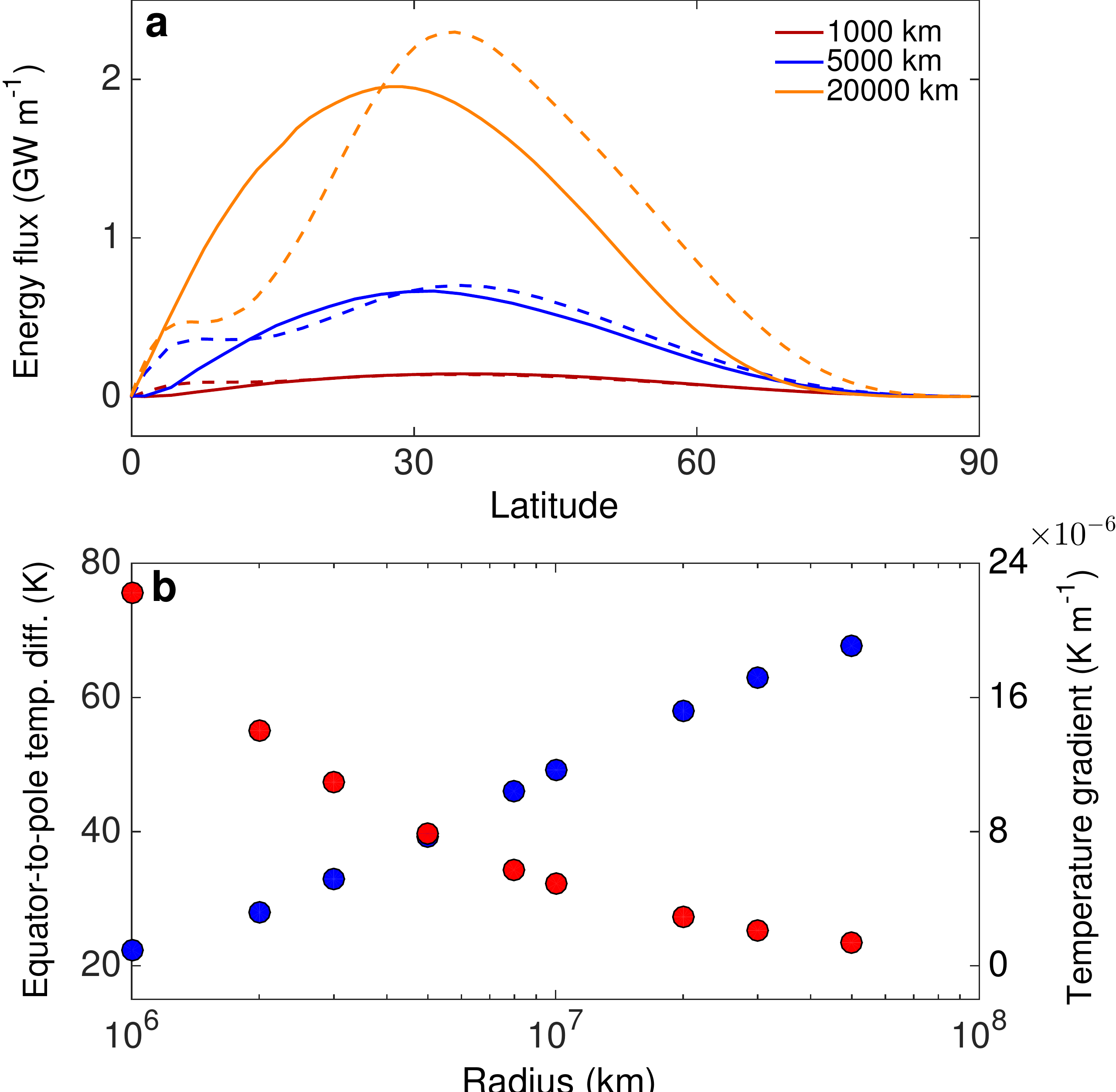}
\par\end{centering}

\caption{\label{fig:delT_radius} (a) The eddy moist (solid) and dry (dashed)
static energy fluxes as function of latitude for three cases with different
planetary radius of 1000~km, 5000~km and 20000~km (b) The equator-to-pole
surface temperature difference (blue) and mean equator-to-pole surface
temperature gradient (red) as function of the planetary radius. }
\end{figure}

\subsection{Dependence on radius\label{sub:Dependence-on-radius}}

In this set of experiments we keep the planetary mean density constant
at the density of Earth (5520~kg~m$^{-3}$), and vary the radius
and correspondingly the mass and surface gravity of the planet (along
the ``rock'' line in Fig.~\ref{fig:grav_RM}). If gravity would
not have been changing then reducing the radius of the planet has
a similar effect to increasing the rotation rate, since the Rossby
number becomes smaller ($u/\Omega L$), where $L$ is the typical
length scale. However, for a rocky planet changing the radius, results
in a change of mass and surface gravity. Again, in this experiment
we keep the mass of the atmosphere (per unit area) constant by consistently
varying the surface pressure. Fig.~\ref{fig:delT_radius} shows that
larger planets have a smaller mean equator-to-pole temperature gradient
due to the MSE transport increasing with planet size (in Fig.~\ref{fig:delT_radius}a
MSE flux is normalized by planetary radius). For all simulations the MSE
fluxes, in similar to the reference case, are dominated by the eddy
fluxes, with the mean contributing negatively in midlatitude (as in Fig.~\ref{fig:mse_flux}). 
Similar to increasing the rotation rate of the planet (section~\ref{sub:Dependence-on-rotation}), as the planet size is increased, the typical eddy length scale becomes
smaller compared to the size of the planet and therefore is less efficient
in heat transport. This results in a larger equator-to-pole temperature
difference (Fig.~\ref{fig:delT_radius}b.

\section{Discussion and conclusion\label{sec:Discussion-and-conclusion}}

To date, nearly a hundred terrestrial exoplanets have been identified,
and this number is expected to grow significantly over the next few
years as more Earth-sized and sub-Earth-sized planets are
discovered. These planets span a wide range of orbital parameters and
incident stellar fluxes, and they likely also span a wide range of
climatic regimes. In this work, we have attempted to characterize the
basic features of the general circulation of these atmospheres,
focusing on planets that are far enough away from their parent star
so that they are not tidally locked, and therefore to leading order,
like Earth, have a zonally symmetric climate. In this sense this study
is different than many recent studies focusing on the atmospheric
dynamics of tidally locked terrestrial exoplanets (e.g.,
\citealp{Joshi1997,Joshi2003,Merlis2010,Heng2011,Wordsworth2011,
Selsis2011, Yang2013, Yang2014, Hu2014, Heng2015, Showman2015}). For
simplicity we have attempted to keep the model configuration as simple
as possible (aquaplanet, no seasons, no ice, simplified radiation),
and we presented several series of numerical GCM simulations
experiments exploring one parameter at a time.  Being one of the first
studies that are addressing such planetary regimes, we have attempted
to address only the basic features of these diverse planetary
atmospheres. We therefore focus on the equator-to-pole temperature
difference, and the size and intensity of the Hadley cells, Ferrel
cells and extratropical jet streams. We focus on six unique systematic
experiments which contain in our view some of the more interesting
characteristics of the planetary atmospheres: rotation rate, stellar flux, atmospheric mass, surface gravity, optical
thickness and planetary radius. Of course, these six encompass only a
small subset of possible explorable parameters.  An alternate approach
would be to nondimensionalize the equations and vary only the
nondimensional parameters. However, in order to give better intuition
we decided to present our integrations using the real physical
parameters.

The key results for the six sets of simulations presented in section 3
are given below: 
\begin{itemize}
\item Planets with faster rotation rates are characterized by smaller
and weaker Hadley cells, and smaller eddy length scales. This results
in a larger equator-to-pole temperature difference, and weaker
jets. The number of jets grows with rotation rate. Slowly rotating planets will have an equatorial eddy
momentum flux convergence resulting in superrotation (e.g.,
Venus, Titan).
\item The stellar flux has a nonmonotonic response in most
aspects we have examined, due to the strong nonlinear dependence of
water vapor abundance on temperature. Warmer and closer planets will
have smaller equator-to-pole temperature differences due to enhanced
eddy MSE transport due to the latent heat component, which increases
significantly
with temperature. However, planets which are far enough from the their
parent star, so that for their abundance of water vapor the atmospheres
are dry, will also have smaller equator-to-pole temperature differences
due to the overall lower temperatures and larger radiative time
scales. Hadley and Ferrel cells exhibit the same nonmonotonic behavior.
\item Planets with larger atmospheric masses, generally have larger
horizontal fluxes but lower vertical fluxes, resulting in reduced
equator-to-pole temperature differences, and higher surface temperatures.
Hadley and Ferrel cells increase in strength with atmospheric mass due to the increased
mass transport.
\item Planets with larger mean densities and therefore larger surface gravity
have stronger Hadley and Ferrel cells. For these cases despite the
growth of eddy energy with surface gravity, the main controller of
the extratropical temperature is the strengthening of the Ferrel cells,
and therefore the equator-to-pole temperature difference increases
with mean planetary density.
\item Planets with larger optical thickness are warmer across all latitudes,
and the equator-to-pole temperature difference decreases with the
increase of optical thickness due to enhanced poleward eddy MSE transport
(mainly because of increased latent heat transport) in the warmer climates. 
\item Planets with larger radii and consequently larger gravity show a decrease
in equator-to-pole temperature gradient with increasing radius due
to enhanced poleward eddy MSE transport. However, despite the reduction
in temperature gradients, this effect does not compensate for the increase
in distance between the equator and the pole, and therefore overall
the equator-to-pole temperature difference increases.
\item The dependence of the equator-to-pole temperature difference on 
rotation rate, atmospheric mass, and other parameters implies that
dynamics may exert a significant effect on global-mean climate
feedbacks such as the conditions under which a planet transitions into
a globally frozen, "Snowball Earth" state.  Thus, our results imply
that the dynamics influences planetary habitability, including the 
width of the classical habitable zone.
\end{itemize}

A key result, which has important implication for the habitability
of these planets, is the equator-to-pole temperature difference response
to the variation in orbital and atmospheric parameters. As we have
discussed, large-scale atmospheric turbulence
attempts to homogenize latitudinal temperature differences,
mainly through poleward transport of moist static energy. In the simulations
presented here, we find that the equator-to-pole temperature difference
increases for larger rotation rates, planets with larger surface gravity
(density) and larger radius planets, while the equator-to-pole temperature
difference decreases with atmospheric mass, larger optical thickness
and for cooler or warmer planets (depending on water vapor). Varying
the distance to the parent star has a nonmonotonic response in the
equator-to-pole temperature difference due to the nonlinear dependence
of water vapor on temperature. Despite the fact that these results
have been obtained in respect to an Earth like reference atmosphere,
the combination of experiments represent the general trends we expect
to find in any atmosphere. Known examples which are consistent with
our results are the terrestrial type atmospheres of Venus, Mars and
Titan. The slow rotating Venus has a very small equator-to-pole temperature
difference and strong jets, despite having a massive atmosphere (92
bars). Titan has a large global Hadley cells due to its slow rotation, and Mars has a similar
rotation rate to Earth but an atmosphere of only 0.006~bar and therefore
has large equator-to-pole temperature differences. 

Moreover, this analysis has implications for the understanding of 
solar system planetary atmospheres as well. The observed
equator-to-pole temperature difference on Venus is only a few degrees
Kelvin \citep{Prinn1987}, and it is unclear if this is because of Venus's slow
rotation or its massive atmosphere with a resulting long radiative
time scale. Extrapolating from our results for
varying atmospheric mass and rotation rate, which both give smaller
equator-to-pole temperature differences for a 92 bar atmosphere
rotating at $\sim0.004\Omega_e$, respectively
(Figs.~\ref{fig:delT_rotation}a,~\ref{fig:delT_ps}b), implies that both
slow rotation and a massive atmosphere are necessary for such a small
equator-to-pole temperature difference.

As more detailed exoplanet observations become available, we will be able to
provide more constraints to these planetary atmospheres to better
constrain the models. This paper provides a first attempt 
to characterize these atmospheres, focusing on the main drivers of
the atmospheric circulation and the resulting climate. Developing
this mechanistic understanding of what controls the climate, will
allow to better provide constraints on what influences habitability
on these planets.

\acknowledgments
We thank Rei Chemke, Eli Galanti and the reviewer for helpful
comments on this work. YK acknowledges support from the Israeli
Science Foundation (grants 1310/12 and 1859/12), the German-Israeli
Foundation for Scientific Research (grant 2300-2295.14/2011), an
EU-FP7 Marie Curie
Career Integration Grant (CIG-304202) and the Helen Kimmel Center for
Planetary Sciences at the Weizmann Institute of Science. APS acknowledges support from NASA
Origins grant NNX12AI79G.

~
\bibliographystyle{Apj}
\bibliography{yohaisbib}
%\bibliography{/Users/yohai/Box/work/bib/yohaisbib}
%{\footnotesize{\bibliographystyle{/Users/yohai/work/exop/ApJ}
%\bibliography{/Users/yohai/work/bib/yohaisbib}
%}
\end{document}